\documentclass[fleqn,usenatbib]{mnras}

\usepackage{newtxtext,newtxmath}

\usepackage[T1]{fontenc}
\usepackage{ae,aecompl}

\usepackage{afterpage}

 \usepackage{soul}   

\usepackage[table,xcdraw]{xcolor}
\usepackage{graphicx}	
\usepackage{amsmath}	
\usepackage{amssymb}	

\title[Age Gradients throughout the Galaxy]{Age Gradients throughout the Galaxy with O-Miras}

\author[J. Grady et al.]{
J. Grady,$^{1}$\thanks{E-mail: jrg71,vasily,nwe@cam.ac.uk}
V. Belokurov,$^{1,2}$
N.W. Evans$^{1}$
\\
$^{1}$Institute of Astronomy, University of Cambridge, Madingley Road, Cambridge, CB3 0HA, United Kingdom\\
$^{2}$Center for Computational Astrophysics, Flatiron Institute, 162 5th Avenue, New York, NY 10010, USA\\
}

\date{Accepted XXX. Received YYY; in original form ZZZ}

\pubyear{2015}

\hypersetup{draft}
\begin{document}
\label{firstpage}
\pagerange{\pageref{firstpage}--\pageref{lastpage}}
\maketitle

\begin{abstract}
We assemble the largest sample of oxygen rich Miras to date and
highlight their importance for age-dating the components of the
Galaxy. Using data from the Catalina Surveys and the
All Sky Automated Survey for Supernovae, we extract a clean sample of
$\sim 2,400$ O-Miras, stretching from the Galactic Bulge to the distant
halo. Given that the period of O-Miras correlates with age, this
offers a new way of determining age gradients throughout the
Galaxy. We use our sample to show (i) disc O-Miras have periods
increasing on moving outwards from $\sim $ 3 to 15 kpc, so the outer
disc O-Miras are younger than the inner disc, (ii) the transition from
younger disc to halo O-Miras occurs at $r \sim 15$ kpc and is marked
by a plummeting in period, (iii) there exists a population of young
O-Miras likely kicked from the disc to heights of order of
$|Z|\sim10$\,kpc, (iv) great circle counts of old Miras show strong
evidence for distant debris agglomeration associated with the
Magellanic Clouds, (v) seven stars in our samples are located at
distances between 200 and 500 kpc surpassing all previously
established records, and, finally, (vi) O-Miras may be present in the
Fornax, Sculptor, Sextans and Leo II Galactic dwarf spheroidals, as
well as the distant globular cluster Pal 4. We spotlight the
importance of O-Mira in the Era of Gaia as universal chronometers of
the Galactic populations.
\end{abstract}




\section{Introduction}

Miras are variable Asymptotic Giant Branch (AGB) stars. They are in the late stages of evolution and will expel their envelopes to form planetary nebulae in a few million years. Typically, they have periods longer than 100 days, and amplitudes greater than 2.5 magnitudes at visual wavelengths~\citep{Ha03}. 

They are worthy of special study for a number of reasons. Firstly, they are the most luminous representatives of the intermediate-age population. Miras are so intrinsically bright that they are detectable throughout the Galactic disc, bulge and halo~\citep{Ha03}, the Magellanic Clouds~\citep{De17}, the dwarf satellites~\citep{Wh09,Sa12} and even in M31 and M33~\citep{An04}. Secondly, they provide an independent distance calibration via period--luminosity relations~\citep{Wh08}. Both oxygen rich (O-Miras) and carbon rich (C-Miras) obey separate period--luminosity-colour relations~\citep{Fe89}. For O-Miras, the colour term in the period-luminosity relation is a very minor correction~\citep{Hu18}. Thirdly, they are major contributors to the processed material currently entering the interstellar medium and an important source of enrichment. However, perhaps the most important property of Miras is that they provide good chronometers~\citep{Fe06,Fe09}. 

Accurate stellar ages have the potential to completely transform Galactic astronomy. The dissection of the Galaxy according to age would allow us to map its assembly and growth, as well as to study secular processes that control its evolution. Unhappily, good stellar ages are unusually hard to come by! The most successful attempts so far use a combination of spectroscopy and stellar evolutionary models. For example, the masses of red giants can be predicted from their photospheric carbon and nitrogen abundances, together with their spectroscopic stellar parameters. With the masses in hand, stellar evolutionary theory is then used to obtain ages, typically with errors of $\sim$ 40 per cent~\citep[e.g.,][]{Ma16,Ne16}.  

Variable stars also offer grand opportunities to derive stellar ages. For example, the period of a Cepheid variable is dependent on its luminosity. An increase in luminosity implies an increase in the stellar mass and a decrease in the Cepheid age. This can be made formal through period-age or period-age-colour relationships, which depend on whether the Cepheid is pulsating in the fundamental or first overtone mode~\citep{Ef78,Bo05}. This can give Cepheid ages with errors of $\sim$ 20 per cent. As Cepheids are tracers of young stellar populations, this method is used to date the age distribution and star formation history in the LMC disc. In principle, age estimates from variable stars depend primarily on the period, which is unaffected by systematics, such as reddening or distance errors. Even relative age estimates for an abundant variable population can supply constraints on the existence of age gradients.

It has long been known empirically that O-Mira kinematics depend on their periods. For nearby O-Miras, the age and velocity dispersion relationship calibrated for stars in the solar neighborhood can be used to derive a period-age relationship~\citep{Fe00,Fe07}. The period of a Mira increases with increasing luminosity, and hence mass. Thus,
longer period Miras correspond to younger ages. Samples of O-rich Miras can therefore be used to determine age gradients in the Milky Way Galaxy and even, in the most propitious circumstances, date the populations. Additionally, the measured spread in near-infrared colours of O-rich Miras is substantially smaller than that observed in the C-rich counterparts \citep[e.g.][]{Whitelock2006,Matsuura2009,Yu17}. One possible explanation of this phenomenon is that the O-Miras suffer less from varying levels of extinction due to circumstellar dust. If true, this makes them more trustworthy distance indicators than their C-rich peers.

Here, we exploit two new datasets to build the largest sample of O-Miras so far assembled. It contains $\sim 2,400$ O-Miras with remarkable radial coverage, Galactocentric radii between 1 and 200 kpc. We use it to identify age gradients throughout the Galaxy, to study the transition in ages between disc and halo O-Miras and to compare the relative ages of different Galactic populations as a function of Galactocentric distance. 

The paper is arranged as follows. Section~\ref{sec:data} presents the Mira dataset, the selection of the oxygen-rich sample, as well as the calculation and validation of the Mira distances. Section~\ref{sec:res} details the application of O-rich Miras to studies of population ages in the Galactic disc and halo, the identification of substructure at large distances and the populations of the dwarf spheroidal galaxies. Section 4 summarises with a discussion of future prospects as the Gaia Era enters high noon.

\begin{figure}
\centering
	\includegraphics[width=\columnwidth]{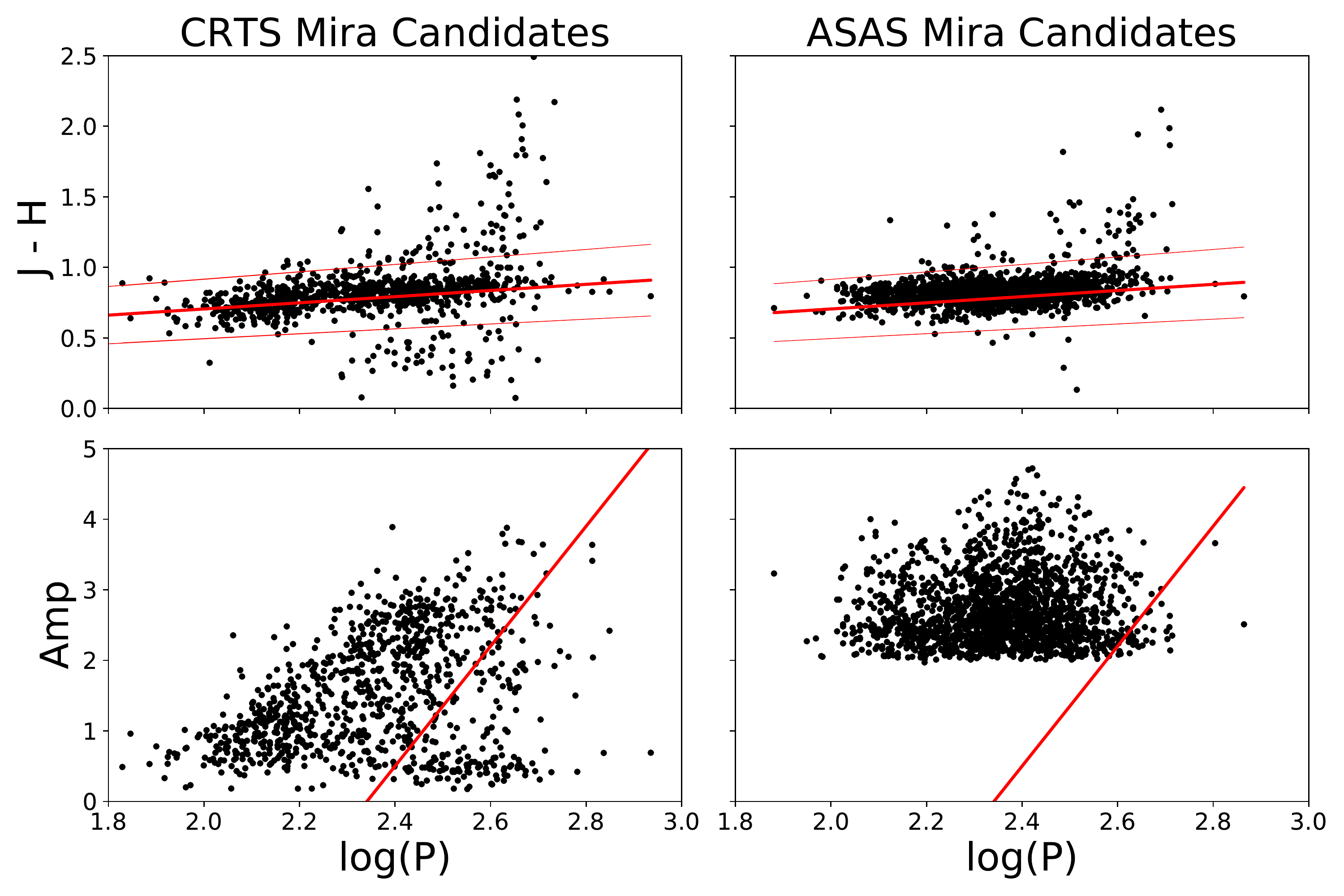}
    \caption[width=\columnwidth]{Red lines show the O-Mira selection cuts applied to our sample. Thin red lines in the upper panels are the 1-$\sigma$ bounds on the cut in colour-period space. Notice that the sequence of O-Miras is only weakly dependent on colour. The red line in the lower panel trims the sample by removing  some contaminating C-Miras using a period-amplitude cut. The left column shows the CRTS data, the right column the ASAS data. }
\label{fig:cuts}
\end{figure}

\begin{figure}
\centering
	\includegraphics[width=\columnwidth]{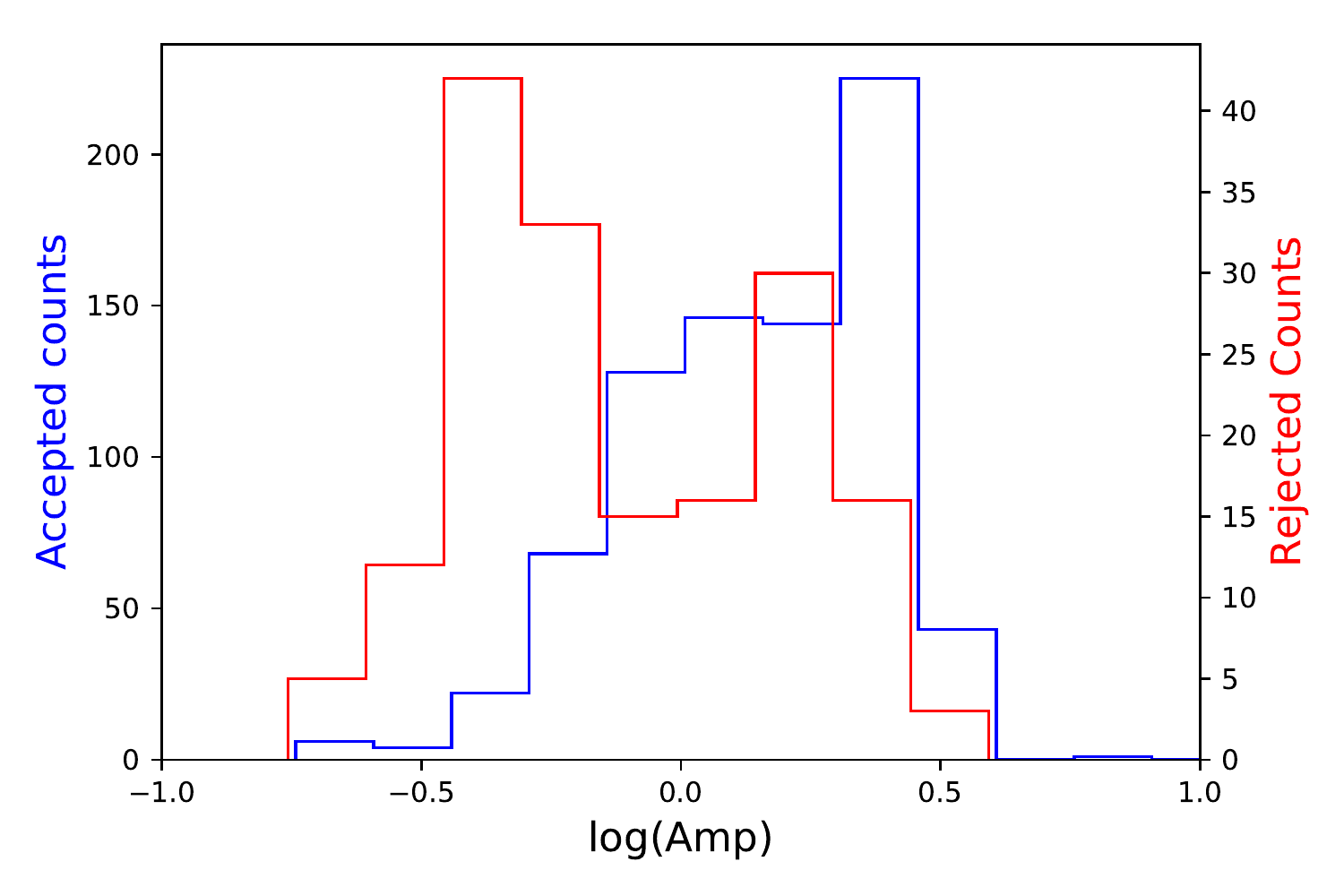}
    \caption[width=\columnwidth]{Amplitude histograms for O-Mira candidates that were accepted (blue) and rejected (red) from the amplitude cut of Fig.~\ref{fig:cuts}. The bimodality of the red histogram is expected, as the contaminants are mainly from two disparate populations, the C-Miras and the semi-regular variables. Bins of width 0.1 were implemented.}
\label{fig:amp_hist}
\end{figure}
\begin{figure*}
\centering
	\includegraphics[width=\textwidth]{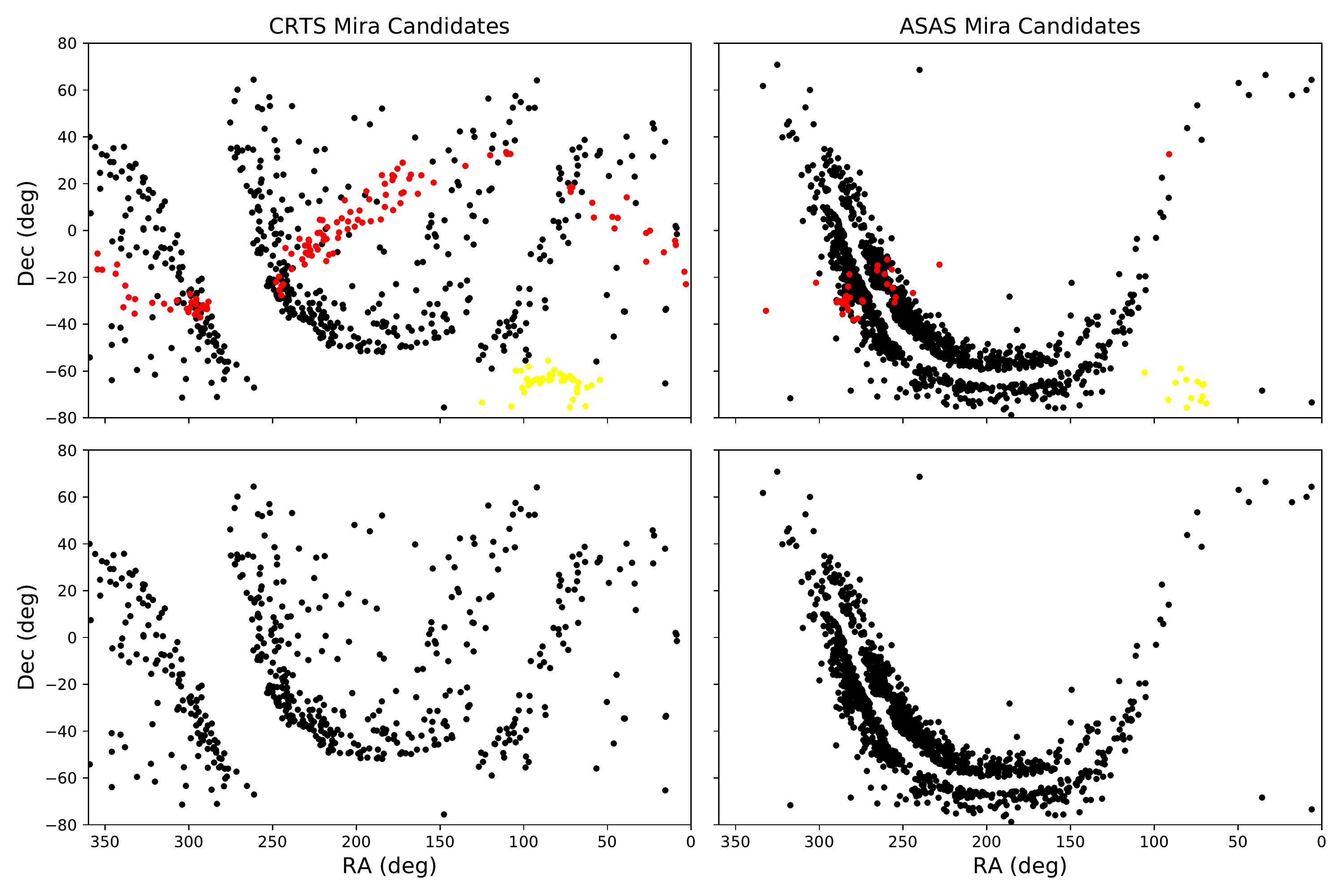}
    \caption{Spatial distributions of selected O-Miras with (upper panel) and without (lower panel) prominent substructures in right ascension and declination for the CRTS (left) and ASAS (right) surveys. Red and yellow markers identify O-Miras associated with the Sagittarius Stream and the Large Magellanic Cloud respectively. Notice that ASAS sample is restricted mainly to the disc, but CRTS samples the halo as well. The percentage of Mira selected to inhabit the LMC and Sgr Stream are $\sim$ 2$\%$ and 6$\%$ respectively.}
\label{fig:ra_dec}
\end{figure*}

\section{Data}
\label{sec:data}

This work exploits data from two primary sources, that of the Catalina Surveys and the All Sky Automated Survey for Supernovae (ASAS). The Catalina Surveys catalog is comprised of two main components surveying the Northern~\citep{crts_north} and Southern~\citep{crts_south} sky respectively. Both surveys are analysed by the Catalina Real-Time Transient Survey (CRTS) in search of optical transient phenomena. Data is extracted from both subsets and will referred to as CRTS hereon. Our sample of Mira candidates is then cross-matched with the 2MASS catalog to obtain ${JHK}_{s}$ magnitudes, yielding a total of 960 unique sources. The ASAS project is an all-sky optical survey with published classifications of variable stars~\citep{ASAS}. Cross-matching the ASAS Mira candidates with 2MASS provided 1,831 sources for study. A cross-match radius of 1 arcesec was used in both instances. Magnitudes are dereddened with extinction coefficients of $\mathcal{R}_{J}$ = 0.72, $\mathcal{R}_{H}$ = 0.46 and $\mathcal{R}_{K_{s}}$ = 0.306 combined with the reddening values of \citet{schlegel}.

\subsection{Mira Selection}
\label{sec:cuts} 

C-Miras (and their allies, the Carbon stars) have been studied extensively before~\citep[e.g.,][]{Do04,Ba14}. They were exploited most recently by \citet{Hu15} to trace out structures in the stellar halo. In this study, we will focus on the less well-studied O-Miras.

The selection of a clean sample of O-Miras proceeds via the application of two main cuts. The work of \citet{Yu17} and \citet{glass_orich_line} shows that the colour indices of Mira increase as a function of period, itself a proxy for age, with linear relations provided by \citet{glass_orich_line} for O-Miras in the SAAO photometric system. For this work, the $J-H$ relation is converted into the 2MASS photometric system via the transformation of \citet{colour_transform} and is shown as the solid red line in the top two panels of Fig.~\ref{fig:cuts}. Associated errors of the original linear relations and colour transformations are combined to produce a 1-$\sigma$ bound defining the selection cut for O-Mira form our sample. These are displayed as the thin red lines in the upper panels of Fig.~\ref{fig:cuts}. Notice that the O-Mira form a tight sequence with only a mild dependence on colour in these plots.

To further rid our sample of C-Miras and other variable contaminants, a cut in amplitude in the $V$-band is applied. The work of \citet{ogle_lpv_lmc} reveals a distinction between O-rich and C-rich stars in amplitude-period space, on which we base our cut shown in the bottom row of Fig.~\ref{fig:cuts}. The CRTS catalog groups Mira and semi-regular variables (SRV), which are themselves C-rich, under the single classification of long period variables (LPV). Accordingly, we use our CRTS sample to define the cut in amplitude based on the higher potential for contamination and the fact that the ASAS sample is predominantly an O-rich population, as evidenced by the top right panel of Fig.~\ref{fig:cuts}. Once the cut is applied, we expect that the rejected C-rich sources display a bimodal distribution in amplitude space given that they are constituted from two components: C-Miras and SRVs. Fig.~\ref{fig:amp_hist} shows this to be the case. This is also in agreement with the results of \citet{ogle_lpv_lmc} who probed the Mira amplitude distribution in the OGLE $\it{I}$ band.

The combined CRTS and ASAS cleaned samples of O-Mira consist of 2,447 sources or $\sim90\%$ of the original combined sample. We do not study the rejected sample of mainly C-Miras any further in this paper.

\begin{figure}
\centering
\includegraphics[width=\columnwidth]{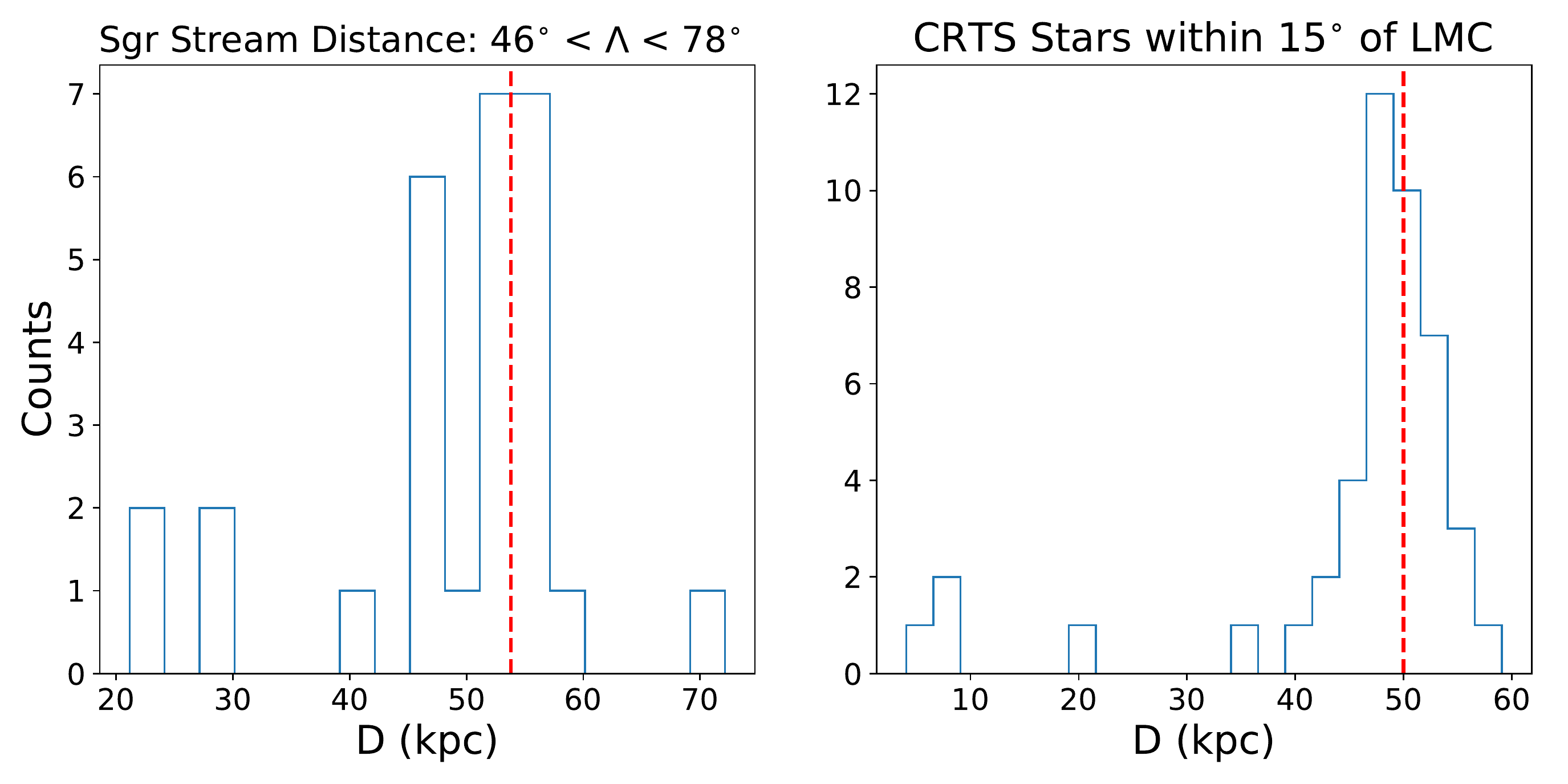}
    \caption{Left panel: Distance histogram of the Sgr Stream selected Mira. The red hashed line indicates the distance to this section of the stream from \citet{vasily_stream}. Right panel: Distance histogram of the LMC Mira. The red hashed line indicates the distance estimate to the LMC from \citet{lmc_dist1}. Bins of width 3 kpc have been used in both plots.}
    \label{fig:dist_hist}
\end{figure}
%


\subsection{O-Mira Distributions}

The spatial distribution of our O-Mira sample is displayed in Fig.~\ref{fig:ra_dec}. The ASAS O-Miras lie close to the Galactic plane with approximately 81$\%$ of the stars having $|b| < 10^{\circ}$. This enables us to probe the O-Mira structure within the Galactic disc. O-Mira from the CRTS selection extend to much higher latitudes and thus grant us information on the halo component of the Galaxy. Of course, the spatial distribution is strongly affected by selection biases of the parent surveys. However, we plan to use the O-Miras to trace age gradients throughout the Galaxy, for which purpose they are unbiased.

We immediately see in Fig.~\ref{fig:ra_dec} a number of prominent substructures, including the Large Magellanic Cloud (LMC) and the Sagittarius Stream (Sgr Stream). They are identified and removed from our sample. Although interesting from the viewpoint of substructure studies, they are a nuisance here as our aim is to understand the relative ages of the mainstream Galactic populations.

We remove sources associated with the LMC by excising a 15$^{\circ}$ radii circle centered on the respective structure. Identifying Sgr Stream denizens requires a transformation into the Sgr Stream coordinates ($L,B$) of ~\citet{vasily_stream}. We remove stars with Sgr Stream latitude $|B|< 10^\circ$ and heliocentric $D > 15$ kpc. The latter is needed as the Sgr Stream passes through the Galactic plane, yet we wish to retain the disc O-Miras. Once removed, our sample of O-Miras is comprised of 2,233 distinct sources, probing the Galactic thin and thick discs, bulge and halo.

\subsection{Distance Determination and Validation}
\label{sec:distance}

We calculate Mira distances using the empirical quadratic Period-Luminosity relation from equation 1 of \citet{Yu17}. We choose to work in the $K_{s}$ band to minimise the effects of extinction. Current best estimates of the distance to the LMC from the work of \citet{lmc_dist1} yield values of 50.3 $\pm$ 0.5 kpc. We then select our LMC Mira candidates (i.e. those within 15$^{\circ}$ of
the dwarf's center), compute their distance and apply a small empirical correction (0.17 mag) to the period luminosity relation to produce a distance distribution peaked at the 50 kpc range. The
distance distribution of these Miras is shown in the right panel of Fig.~\ref{fig:dist_hist}. The median distance of the prominent peak is $49.8$ kpc. The standard deviation approximated from the median absolute deviation of the LMC sample is $\sim 4$\,kpc. This can be used as a typical distance error to O-Mira stars in our sample, which, encouragingly, appears to be $\lesssim10\%$. This method results in a final period luminosity relation of:
\begin{equation}
M_{K_{s}} = -6.90 - 3.77\left ( \textup{log}P -2.3 \right ) -2.33\left ( \textup{log}P -2.3 \right )^{2} -0.17
\end{equation}
where $P$ is the period in days and $M_{K_{s}}$ is the absolute magnitude in the $K_{s}$ band. The logarithm here, and henceforth, is to base 10. As a further check, the distance to a well populated section of the Sgr Stream was computed for the associated O-Mira. The left panel of Fig.~\ref{fig:dist_hist} shows the peak of the distribution to lie at $\sim$ 50 kpc in agreement with the values in Table 1 of \citet{vasily_stream} which quote distances of 45 - 58 kpc in the chosen Sgr Stream longitude range.

\begin{figure}
\centering
	\includegraphics[width=\columnwidth]{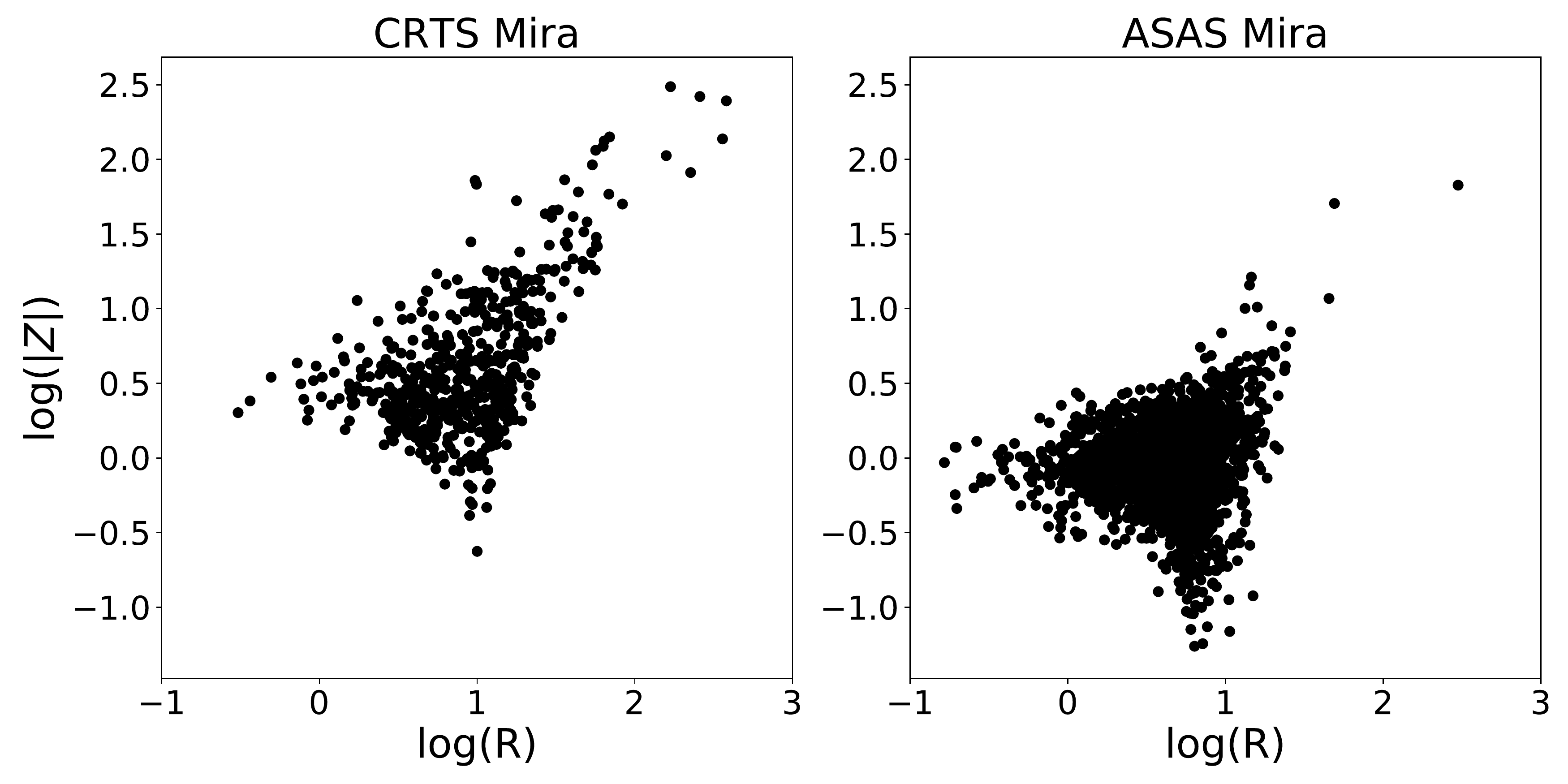}
    \caption[width=\columnwidth]{Locations of O-Miras in CRTS (left) and ASAS (right) using cylindrical radius $R$ and vertical height $|Z|$ in kpc. Notice that the ASAS O-Miras do not extend      beyond $\sim 1$ kpc from the Galactic plane and so are predominantly thin and thick disc stars. The CRTS O-Miras range in radii from 0.1 kpc to more than a hundred kpc, from the Galactic bulge into the distant halo.}
    \label{fig:logR_logZ}
\end{figure}

Taking the solar position as $R_\odot=8$ kpc, we convert from Galactic $l, b$ and heliocentric distance to Galactocentric cylindrical polar coordinates ($R,Z$). The panels of Fig.~\ref{fig:logR_logZ} compare the spatial distributions of O-Mira in the CRTS and ASAS samples. In CRTS, the sample probes distances characteristic of the Galactic bulge right out to the far halo. By contrast, in ASAS, we see a clustering of stars at low $|Z|$ around the solar neighbourhood, along with a sample largely confined to within a kpc of the Galactic plane. The two samples are therefore highly complementary, with CRTS probing from the bulge to the halo, and ASAS concentrating on the thin and thick discs.

\begin{figure}
\centering
	\includegraphics[width=0.49\textwidth]{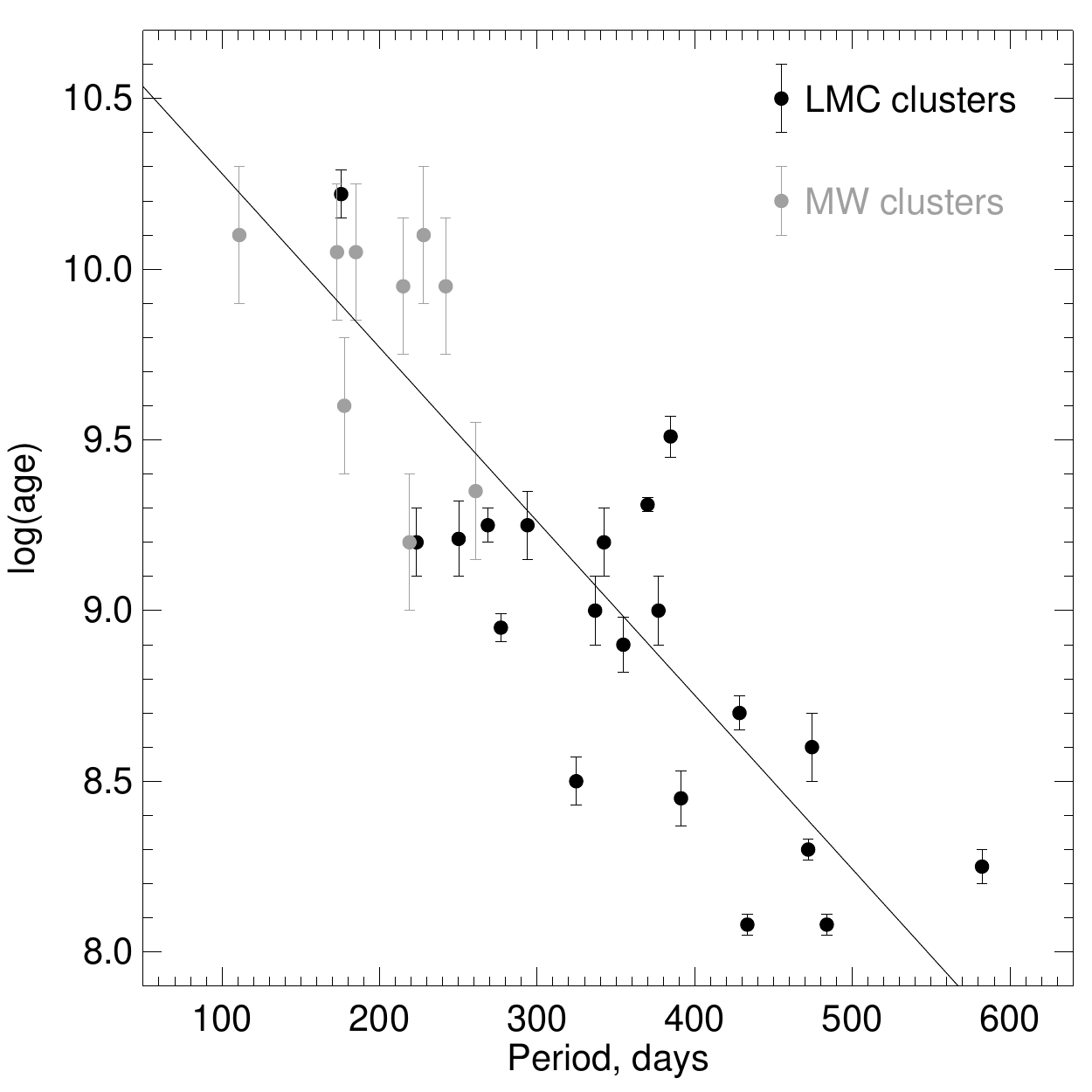}
    \caption[width=\textwidth]{Dependence of the Miras age on its
      period. For the LMC, ages of the massive star clusters from
      \citet{Baumgardt2013} are used by cross-matching their positions with those of the LMC Miras in the catalog by ~\citet{Soszynski2009} with a 100 arcsec aperture. These measurements are complemented by age estimates of the Mira stars in the combined sample of ASAS+CRTS in the vicinity of Galactic star clusters from the catalog of \citet{Kharchenko2016}. The cross-match is carried out with a 9 arcmin aperture, only objects with the distance mismatch less than the half of the distance to the cluster are kept. Straight line indicates a linear fit to the data with 10.79 and $-5.09 \times 10^{-3}$ for the intercept and slope respectively.}
\label{fig:period_age}
\end{figure}

Given the strong dependence of the Miras luminosity on its period, the completeness and contamination of the O-Mira selection must vary with distance. This is dictated by the distribution of the LPV stars in the space spanned by period, colour and amplitude as shown in Fig.~\ref{fig:cuts}. For example, the contamination from C-rich Mira stars and other LPVs with broader $J-H$ colour range increases sharply with period. Thus, because at fixed limiting magnitude the most distant stars detectable are those with larger periods, it is possible to expect that contamination may increase with distance. On the other hand, the short-period O-Mira stars have lower amplitudes and thus may suffer a much faster drop in completeness with distance compared to the long-period counterparts.

\section{Results}
\label{sec:res}

\begin{figure}
\centering
	\includegraphics[width=0.48\textwidth]{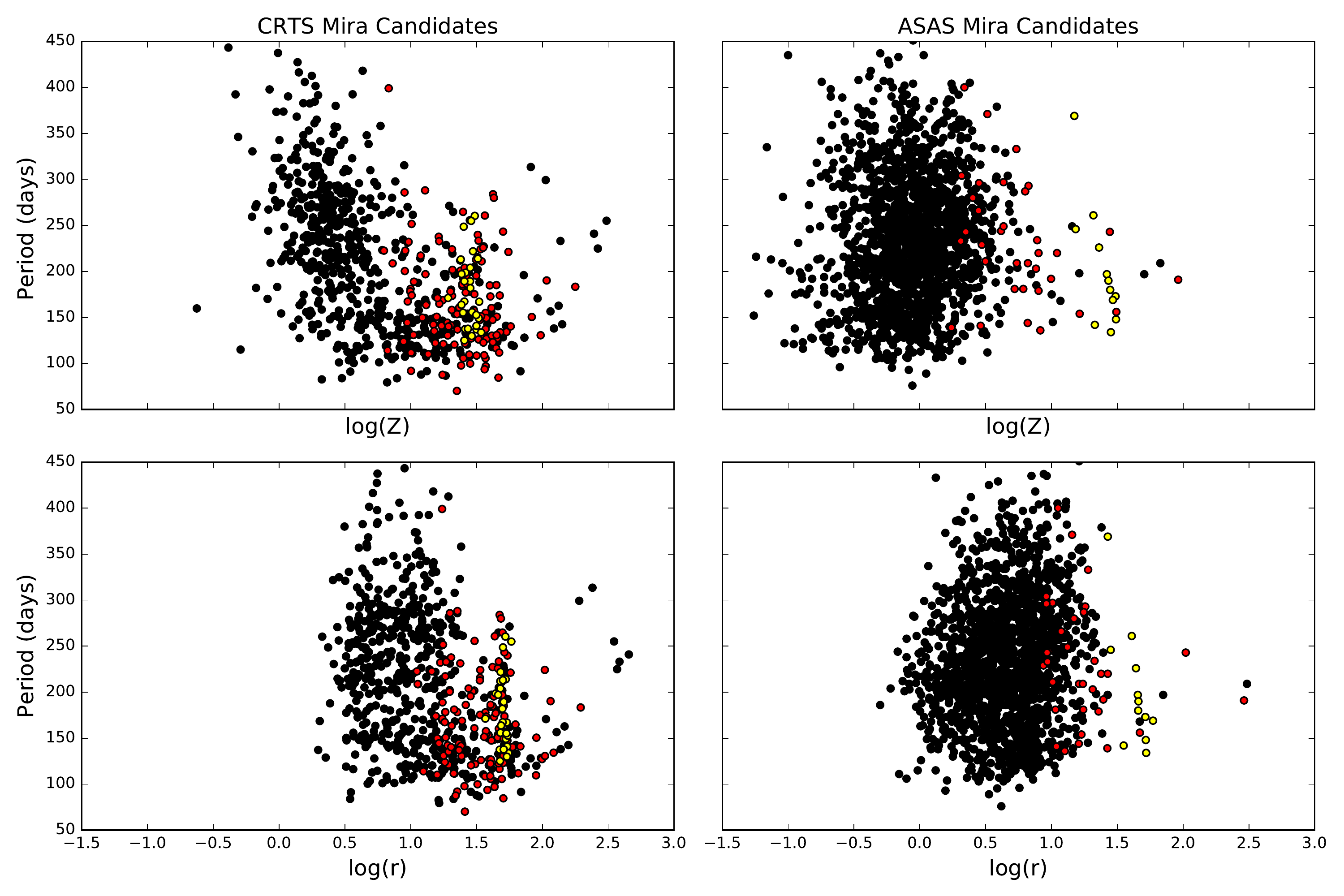}
    \includegraphics[width=0.48\textwidth]{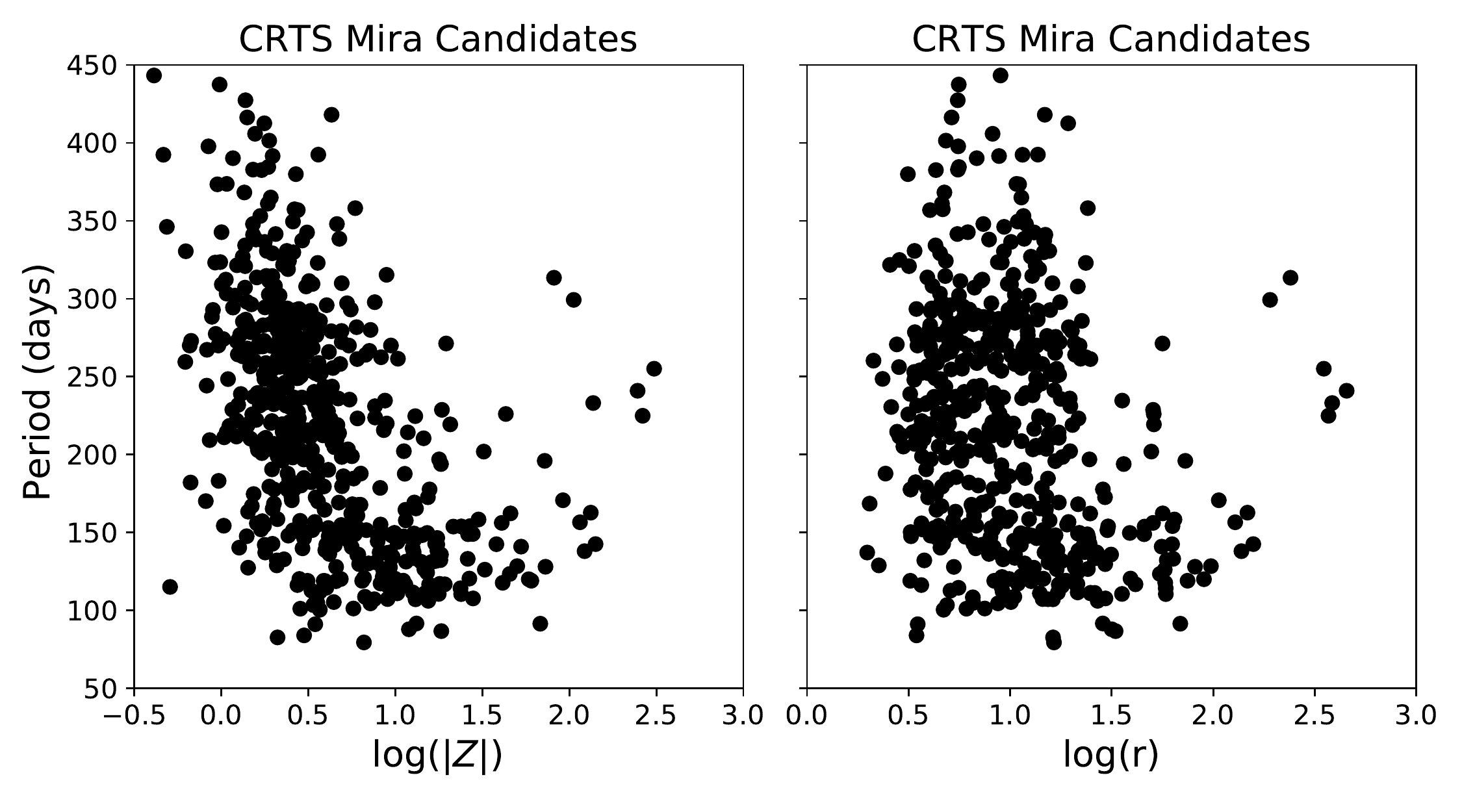}
    \caption[width=\textwidth]{The upper set of four panels display the distribution of O-Mira in period versus Galactic position with the main halo substructures. Red indicates Sgr Stream O-Mira, yellow those of the LMC. Vertical height above/below the plane and spherical polar radius are represented by $|Z|$ and $r$ respectively. The lower two panels display the same distribution for the CRTS Mira, but with the removal of the LMC and Sgr Stream. }
    \label{fig:per_R_Z_scatter_with_struc}
\end{figure}
\begin{figure*}
\centering
	\includegraphics[width=\textwidth]{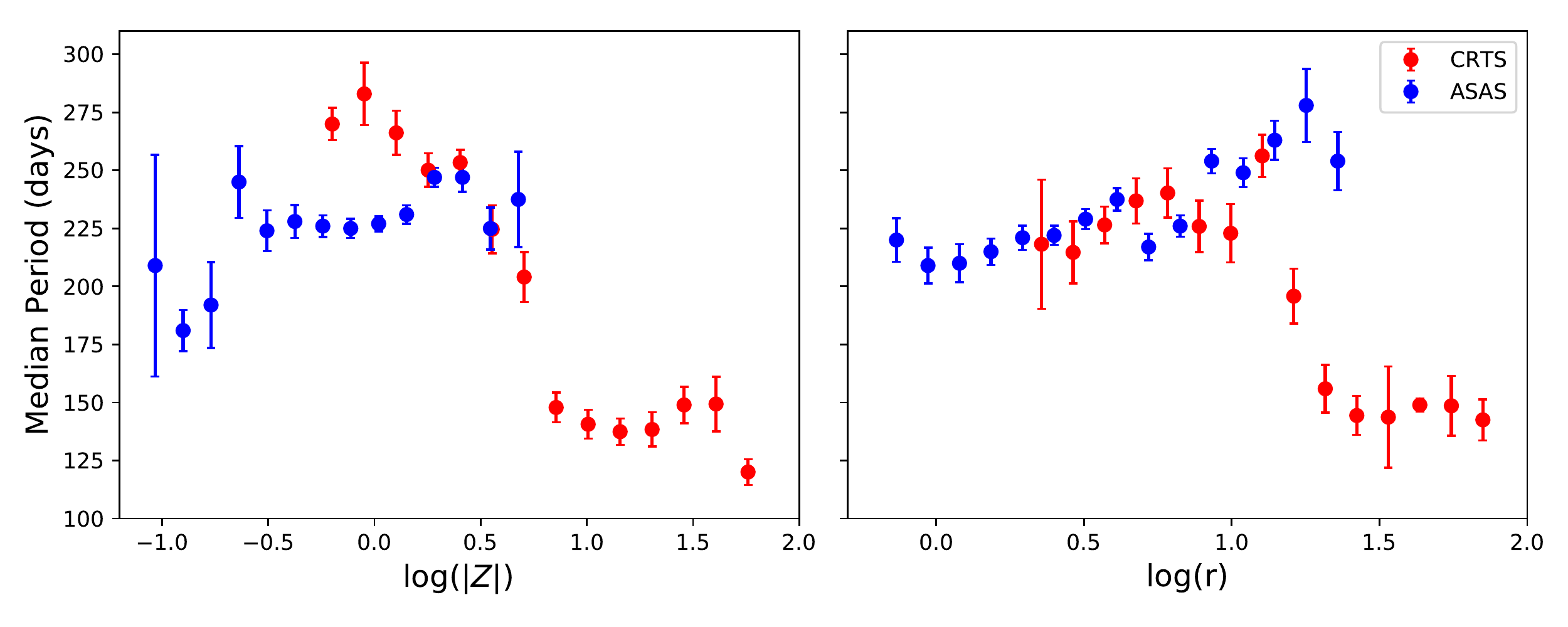}
    \caption[width=\textwidth]{The median of O-Mira periods versus height above/below the Galactic plane $|Z|$ and spherical polar radius $r$ in kpc. Distances have been binned with bins containing less than 5 O-Mira omitted. Error bars indicate standard errors for each bin. Red and blue markers identify CRTS and ASAS Mira respectively.  Notice that the transition from disc to halo O-Miras is marked by a sharp drop in period at $r \sim 15$ kpc in the CRTS sample. The disc O-Miras in ASAS have periods that increase on moving outwards in radii from $\sim 3$ to $15$ kpc, as well as on moving further away from the Galactic plane in height. }
    \label{fig:binned_plots}
\end{figure*}

\subsection{Age Gradients in the Galactic Disc}
\label{sec:age}
%

While the age of the Mira variables has typically been estimated by
matching them to disc stellar populations with similar velocity
dispersions \citep[see e.g.][]{Fe07, Fe09,mira_age}, we attempt to
calibrate the Mira age-period relation by finding candidate LPVs in
the LMC and the Milky Way star clusters with known ages. In the LMC,
we rely on the Mira catalog by \citet{Soszynski2009} and the
compendium of massive star clusters by \citet{Baumgardt2013}. In the
Galaxy, we use the combined ASAS+CRTS sample of O-rich Mira presented
in this work and the catalog of star clusters
by~\citet{Kharchenko2016}. Fig.~\ref{fig:period_age} shows the age of
the star cluster as a function of the period of the candidate Mira
matched to its location. In agreement with previous similar studies
\citep[][]{Nishida2000,Kamath2010}, the trend is clear albeit with
considerable scatter. The shortest period Miras with $P<200$ days do
tend to live in clusters that are 8-10 Gyr old. On the other hand,
Mira stars with periods of $P\sim350$ days are found in much younger
star cluster with ages of order of 1 Gyr.

\begin{figure*}
\centering
	\includegraphics[width=\textwidth]{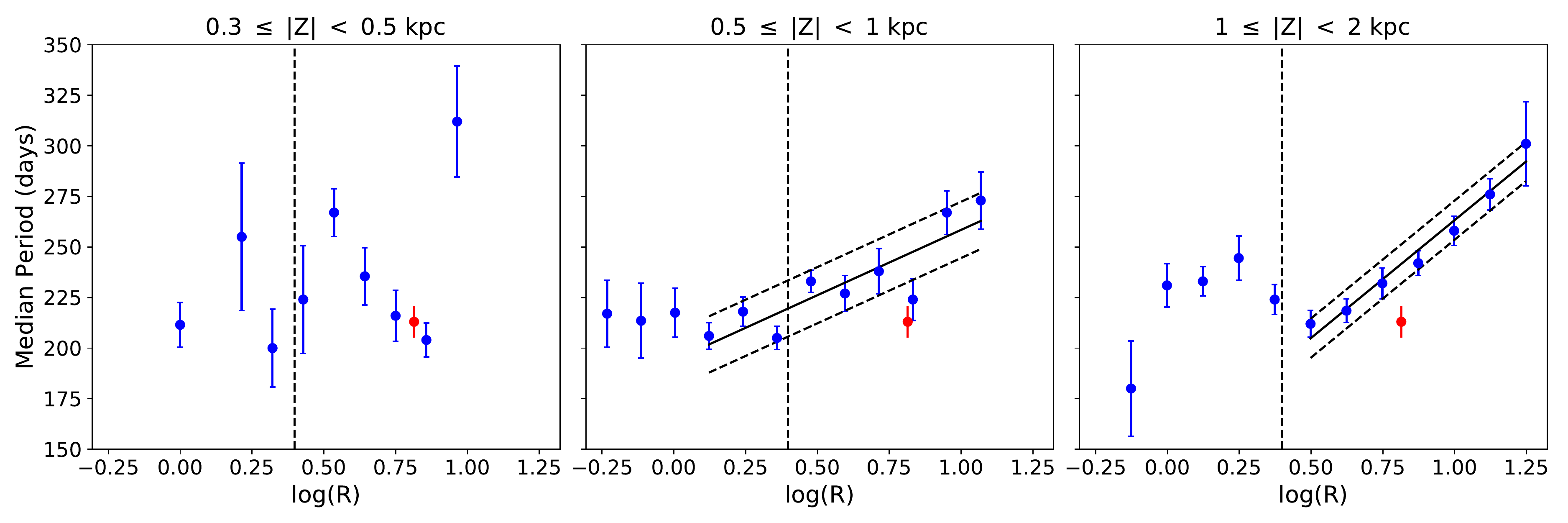}
    \caption[width=\textwidth]{Relation between periods of ASAS O-rich Miras and their distances above and within the plane with cuts in vertical distance. Red points mark solar neighborhood Mira selected with the criteria of log($|Z|$) < -0.5. This selection was chosen given the distribution of ASAS Mira in Fig.~\ref{fig:logR_logZ}. The vertical black dashed line, at $R=2.5$ kpc, marks the boundary between left lying bulge/bar regions and the disc to the right. Bins with less than 5 Mira are neglected.}
    \label{fig:zcuts}
\end{figure*}

\begin{figure*}
\centering
	\includegraphics[width=\textwidth]{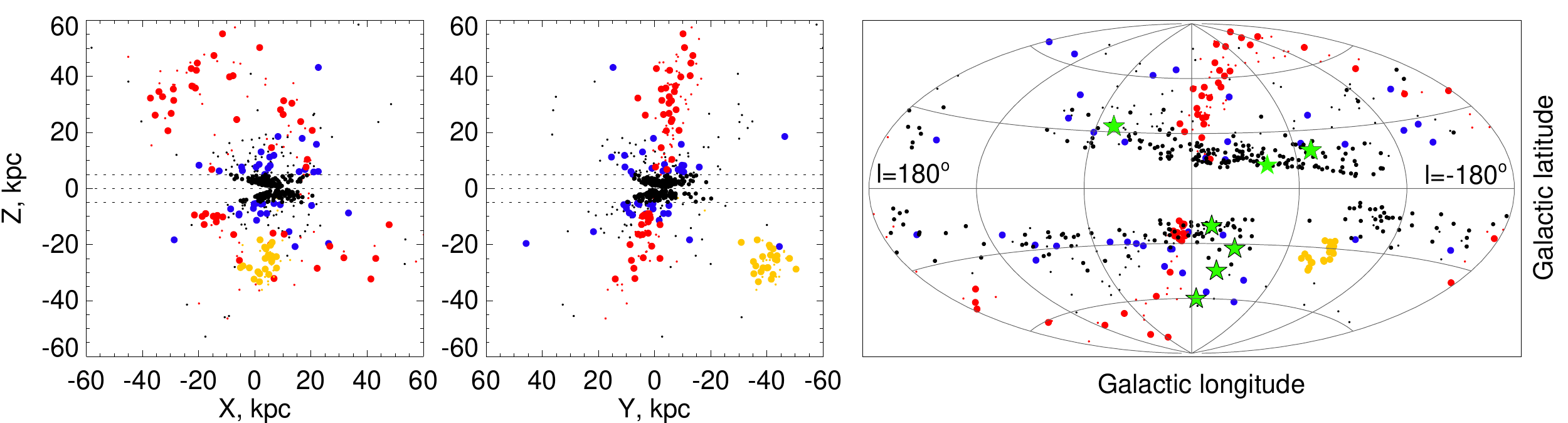}
    \caption[width=\textwidth]{Distribution of long-period (likely
      young) CRTS Mira stars in the Galactic halo. Small dots
      represent Miras with $P<170$d, large filled circles correspond
      to those with $P>170$d. Red symbols are stars projected to lie
      within the Sgr stream and more than 15 kpc away from the
      Galactic center. Yellow symbols mark the stars within
      15$^{\circ}$ of the LMC. Blue (black) filled circles give the
      locations of long-period Mira stars above (below) $|Z|=5$
      kpc. Note that the blue symbols trace out a highly flattened,
      disc-like subsample of Miras traveling to Galactic heights
      $5<z<10$ kpc. Outside of this sub-group, the vast majority of
      long-period Miras at high $Z$ belong either to the Sgr stream or
      the LMC. {\it Left:} Distribution of stars in Galactic $X$ and
      $Z$. {\it Middle:} Distribution of stars in Galactic $Y$ and
      $Z$. {\it Right:} Distribution of stars on the sky in Galactic
      coordinates. Green stars mark the location of O-Miras with
      Galactocentric radii in excess of $\sim200$\,kpc. Of these, 4
      green stars with black outlines lie beyond $350$\,kpc from the
      Milky Way.}
    \label{fig:long_period}
\end{figure*}

We plot Galactocentric spherical polar radius $r$ and vertical height
$|Z|$ above or below the Galactic plane as a function of period with a
view to understanding the O-Mira age distribution in both the disc and
halo components of the Galaxy. This is shown with and without
substructures in the four topmost and bottommost two panels of
Fig.~\ref{fig:per_R_Z_scatter_with_struc}. The CRTS panels of these
plots are the most interesting. They show that O-Miras with short
periods ($\lesssim 200$ days) reach much greater heights above the
Galactic plane than O-Miras with long periods ($\gtrsim 300$ days). A
similar pattern is seen in the plot of period versus distance, with
long period O-Miras confined to within the innermost $r \sim 15$
kpc of the Galaxy. Clearly, we are seeing the transition from a
younger disc O-Mira population into an older halo component. The age profile of the halo is rather flat, seemingly at
odds with the recent claims of the negative age gradient in the halo
by \citet{Carollo2016}. Further, the six most distant stars (beyond 200 kpc) seen in the CRTS sample
are clearly young. Less transition is observed in the ASAS counterpart
owing to the limited sampling at high Galactic latitudes in this
dataset.  The LMC and Sgr Stream O-Mira are seen to predominantly
inhabit an intermediate to low period range representative of the
stellar population these structures contain. This is because
of comparatively recent star formation processes, not seen in the halo in general.

The panels of Fig.~\ref{fig:binned_plots} give the period distributions of the O-Miras in vertical distance from the plane $|Z|$ and spherical polar radius $r$. Distances have been binned, with bins containing less than 5 stars being rejected, and prominent substructures removed. Both mean and median periods have been computed for each bin along with standard errors. The standard error on the median is formulated through appropriate scaling of the median absolute deviation. As the ASAS Mira sample contains predominately disc O-Miras, it is interesting to note the positive correlation between Galactocentric radius and period given that period is a proxy for age. The implication here, then, is that the age of the O-Mira populated components of the disc decreases on moving outwards. We do not see any evidence of a sharp upturn in the radial profile of the median age of disc, as conjectured by \citet{Ro08}. Their simulations suggest that the outer disc is deficient in young stars, but is populated by old stars which migrated from the inner disc.  If such an upturn exists, then it must lie beyond $\sim 15$ kpc. In fact, \citet{Am17} have recently argued that the break may be at $16.1 \pm 1.3$ kpc on the basis of population synthesis applied to 2MASS data.

To probe any change in period gradient, the ASAS O-Mira are further binned into three subgroups: $0.3 \leq |Z| < 0.5$ kpc, $0.5 \leq |Z| < 1$ kpc and $1 \leq |Z| < 2$ kpc. We use cylindrical radius $R$ in this instance to focus on the evolution of period throughout the Galactic disc. These cuts correspond to radial ranges of $0.76 \leq R < 14.3$ kpc, $0.25 \leq R < 18.4$ kpc and $0.67 \leq R < 21.7$ kpc respectively, as enforced by the footprint in Fig.~\ref{fig:logR_logZ}. The middle and rightmost panels of Fig.~\ref{fig:zcuts} clearly demonstrate that beyond 2.5 kpc the O-Mira population follows the trend of increasing median period with increasing Galactocentric distance. This can be quantified by linear fits yielding:
\begin{eqnarray}
	\langle P \rangle &=& \left( \: 64.6 \log R + 193.9 \: \right) \pm{14} \nonumber \\
	\langle P \rangle &=& \left ( 116.6 \log R + 146.7 \: \right) \pm{10}
\end{eqnarray}
where $\langle P \rangle$ represents the median period and the fits correspond to the ranges of  $0.5 \leq |Z| < 1$ kpc and $1 <\leq|Z| < 2$ kpc respectively. At small radii ($< 2.5$ kpc), the trend is not present and the relationship between median period and distance is flat. This is possibly indicative of existence of the intermediate age Mira reported towards the bulge in the works of \citet{Catchpole_ages}, \citet{bulgemira_feast} and \citet{bulgemira_mats}, which are difficult to separate from the disc Miras.



Recent estimates, from the works of \citet{thickdsk_ojha}, \citet{thickdisk_juric} and \citet{thickdisk_jaya}, for the scale of the thick disc are based on double exponential models with radial and vertical scale heights in the ranges $h_{R} \sim$ 3.6- 4kpc and $h_{z} \sim$ 0.8-1 kpc. In a broad sense, the expected populations in the consecutive panels of Fig.~\ref{fig:zcuts} would be expected to comprise of thin disc, thick disc and halo stars from the left to right. It is interesting to note that radial age gradients at different $|Z|$ ranges have been observed by \citet{RG_gradient} in a population of APOGEE red clump stars, from $R \geq$ 4kpc onward. They see clear separations in age gradients between the $|Z|$ bins, that is, higher $|Z|$ bins constitute a clearly older population for given radius $R$, whereas the distinctive gradients in our sample are less well separated. However, their results are consistent with ours, as we both detect an older stellar population as we moves outwards in the disc. 

Additionally, the red marker in Fig.~\ref{fig:zcuts} corresponds to
the datum derived from a solar neighbourhood selection of Mira. From
Fig.~\ref{fig:logR_logZ}, there is an appendage of Mira extending to
low $|Z|$ at the Solar position. A cut selecting Mira with $\log |Z| <
-0.5$ has been used to define the O-Miras in the solar region. This
though leads to the surprising result that the solar neighbourhood
O-Miras have median periods lower than the thick disc O-Miras, and
hence are older! Although the origin of this result is not fully
clear, there are indications that our solar neighbourhood sample is
affected by survey systematics. In addition to the variation of the
sample completeness and contamination with period as discussed in
Section~\ref{sec:distance}, other effects such as that linked to
saturation may be taking place. Saturation for ASAS is at $V \sim
11$. Nearby long period O-Miras are not only more luminous, but they
also have higher amplitudes, so they are more likely to be saturated
and therefore not included in the selection. Finally, it is worth
pointing out that, if only a sub-class of Miras is considered - such
as O-rich stars, the period-based chronometry argument might only work
for a limited range of ages. This is because the period-age relation
is normally derived for a combination of both C- and O-rich Miras,
however, their period distributions are rather different. It has been
noticed \citep[see e.g.][]{Cioni2001, Fe89, Lorenz2011} that O-rich
stars are biased towards short periods while C-rich ones tend to have
longer periods on average. Translated into age, these differences
imply that O-rich Mira might significantly under-sample the younger
populations of the Galactic disc.


\subsection{Miras outside of the Galactic disc}

While the ASAS sample is largely limited to $|Z|<5$\,kpc, the
complementary CRTS Miras probe a much larger range of Galactic
heights, $5<|Z|<200$ kpc (and probably beyond). We first focus on the
distribution of long-period pulsators in CRTS. According to
Fig.~\ref{fig:binned_plots}, the bulk of the disc Miras in the ASAS
set have periods in excess of 170 days. Note that many lines of --
albeit indirect -- evidence exist that the long-period stars are
substantially younger than their short-period counterparts (see
Fig.~\ref{fig:period_age} for example). Fig.~\ref{fig:long_period}
gives the view of the distribution of the population of the
long-period (and thus likely young) Mira stars across the Milky
Way. Apart from the obvious disc denizens (filled black circles),
three distinct groups can be identified in the Figure. The first two are
the stars belonging to the Sgr stream (red filled circles) and the LMC
(yellow filled circles).

\begin{figure}
\centering
	\includegraphics[width=\columnwidth]{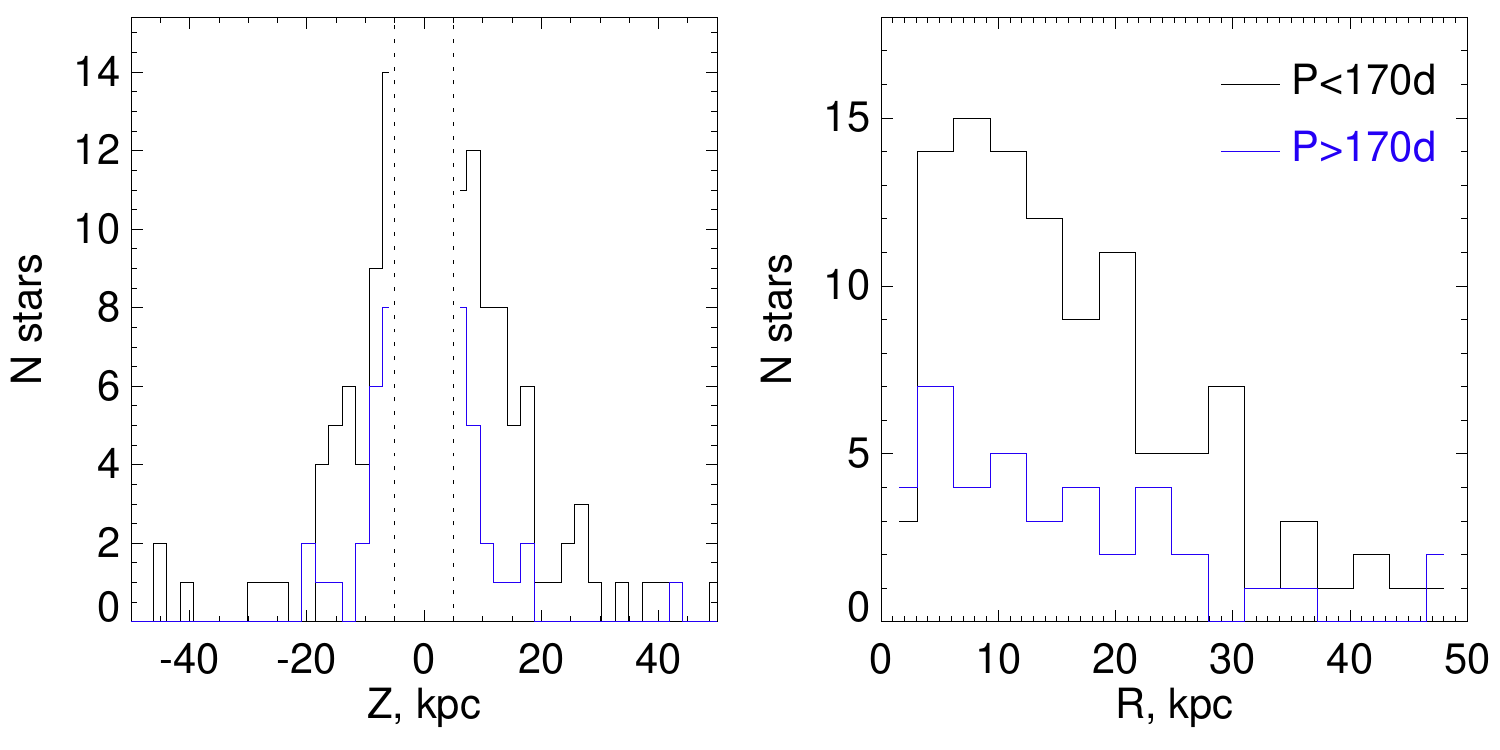}
    \caption[width=\columnwidth]{Difference in spatial extent between
      the old, i.e. short-period (black line) Miras and the young,
      i.e. long-period (blue line) ones as observed by CRTS. Stars
      belong to the Sgr stream and the LMC have been excluded. {\it
        Left:} Distribution of stars as a function of Galactic
      $Z$. Note that the long-period Mira stars (blue line) appear
      to be mostly limited to $|Z|<10$\,kpc. {\it Right:} Distribution
      of stars as a function of Galactocentric cylindrical $R$. Note
      that both young and old, i.e. long- and short-period, stars
      extend as far as $R\sim30$\,kpc.}
    \label{fig:long_period_histo}
\end{figure}

The last group (blue filled circles) is comprised of stars that follow
the extent of the disc in $X,Y$ dimensions but travel to heights as
large as $|Z|\sim10$\,kpc. Their reach in $|Z|$ and $R$ can be gleaned
from Fig.~\ref{fig:long_period_histo}, where their distribution
(blue histograms) is compared to that of the short-period Miras (black
histograms). Indeed, both long-period and short-period objects are
detected as far as $R\sim30$\,kpc, however the (likely) young CRTS
Miras do not wander beyond $|Z|\sim10$\,kpc. We therefore speculate
that, if our distance estimates are valid, these objects represent a
population of kicked disc stars. As of today, several prominent groups
of high $|Z|$ disc stars are known including those composed of M-giant
stars, close relatives of Miras
\citep[][]{RochaPinto2004,Xu2015,Bergemann2018,Sheffield2018,Deason2018,deBoer2018}. Strikingly,
apart from the three obvious groups, such as the kicked disc
population (blue) and the members of the Sgr stream and the LMC (red
and yellow), there are almost no young Miras in the Galactic halo.

However, we do find a total of 7 long-period Miras that go much
beyond the rest of the sample, i.e. to Galactocentric distances in
excess of $\sim200$\,kpc and are not projected behind the Sgr stream
of the LMC on the sky. Six of these are part of the CRTS catalog and
one from ASAS. The locations of these Miras are marked with filled
green stars in the right panel of
Fig.~\ref{fig:long_period}. Moreover, we use black outlines to point
out 4 remarkable objects (all from CRTS) with $r>350$\,kpc. Please
note that at this stage, these 7 objects should be viewed as
candidates. This is because, as explained in
Section~\ref{sec:distance}, the sample contamination likely increases
with period and all of these Miras have periods larger than 220 days,
as presented in Table~\ref{table:distant_mira}. If our absolute
magnitude estimates are correct, then all 4 of the Miras marked with
green stars and black outlines in Fig.~\ref{fig:long_period} lie
further than the previously identified most distant Milky Way
stars. There are however, several reasons to be cautious. As judged by
the contents of Table~\ref{table:distant_mira}, some of the distant
Miras have suspiciously low amplitudes given their relatively long
period. It is of course not impossible that their amplitudes are
under-estimated given that these objects are some of the faintest
Miras in our sample. Another possibility is that these stars are
misclassified and are not O-Miras but rather C-rich LPVs instead. In
that case, we have checked that simply applying the
period-colour-luminosity relation appropriate for C-Miras given in
\citet{Yu17} would yield distances similar to those listed in the
Table or larger. Note that for C-Miras, PCLRs typically yield larger
uncertainties compared to those derived for O-Miras, indicative, most
likely of unknown (and quite often significant) amounts of
circumstellar dust.

\begin{table*}
\centering
\begin{tabular}{|cccccccccc|}
\hline
\rowcolor[HTML]{C0C0C0} 
\textbf{RA (deg)} & \textbf{Dec (deg)} & \textbf{$\ell$} & \textbf{$b$} & \textbf{$J$} & \textbf{$H$} & \textbf{$K_{s}$} & \textbf{Period (days)} & \textbf{Amplitude} & \textbf{D (kpc)} \\
\hline
257.7 & 27.1 & 49.0 & 33.0 & 14.5 & 13.9 & 13.6 & 299 & 1.5 & 191 $\pm$ 15 \\
177.3 & -41.7 & 290.7 & 19.7 & 14.9 & 14.2 & 14.0 & 314 & 1.4 & 240 $\pm$ 20 \\
283.8 & -48.1 & 348.4 & -20.4 & 16.8 & 15.9 & 15.6 & 233 & 1.2 & 386 $\pm$ 31 \\
301.7 & -63.4 & 333.1 & -32.5 & 16.4 & 15.9 & 15.9 & 241 & 1.0 & 454 $\pm$ 36 \\
322.6 & -54.0 & 342.5 & -44.7 & 16.5 & 15.9 & 15.6 & 225 & 1.5 & 369 $\pm$ 30 \\
340.8 & -41.5 & 355.8 & -60.1 & 16.0 & 15.4 & 15.2 & 255 & 0.9 & 350 $\pm$ 28\\
213.9 & -48.1 & 317.2 & 12.4 & 16.2 & 15.4 & 15.3 & 209 & 2.3 & 306 $\pm$ 24 \\
\hline
\end{tabular}
\caption{Table of the seven most distant Mira in our sample. Listed are the heliocentric coordinates, Galactic coordinates, 2MASS magnitudes and periods. We also provide values of the amplitude and Galactocentric distance values. These Mira correspond to the green markers in the rightmost panel of Fig.~\ref{fig:long_period}. Distance errors are estimated from the dispersion of our LMC Mira distribution in Fig.~\ref{fig:dist_hist}.   }
\label{table:distant_mira}
\end{table*}

\begin{figure}
\centering
	\includegraphics[width=\columnwidth]{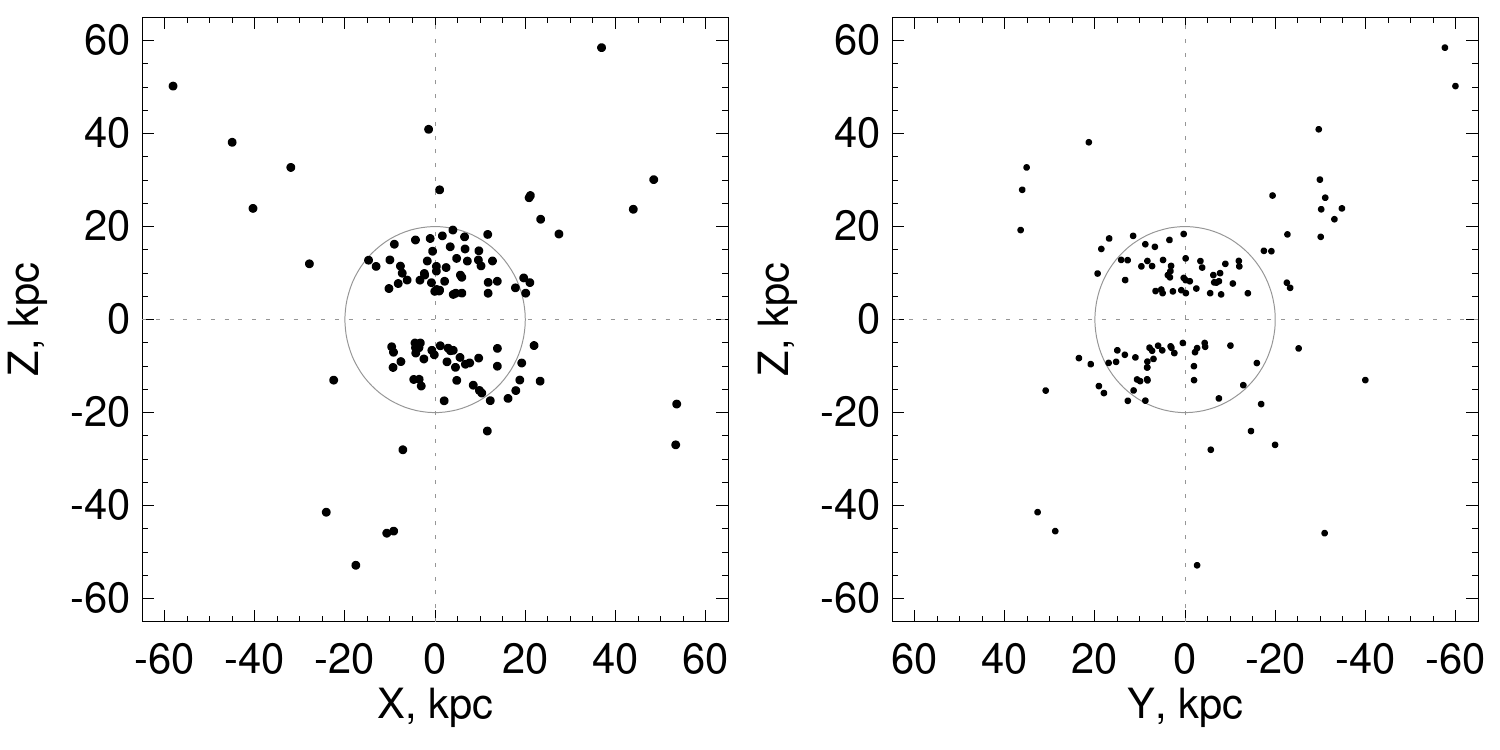}
    \caption[width=\columnwidth]{Distribution of short-period, i.e old
      Mira stars in the Milky Way halo as viewed by CRTS. Stars
      belonging to the Sgr stream, the LMC and those with $|Z|<5$\,kpc
      have been excluded. Note the marked drop in the density of old
      Mira stars outside of the grey circle, which has a radius of 20
      kpc. {\it Left:} Distribution of stars projected onto the $X, Z$
      plane. {\it Right:} Distribution of stars projected onto the $Y,
      Z$ plane.}
    \label{fig:short_period_xyz}
\end{figure}
\begin{figure*}
\centering
	\includegraphics[width=\textwidth]{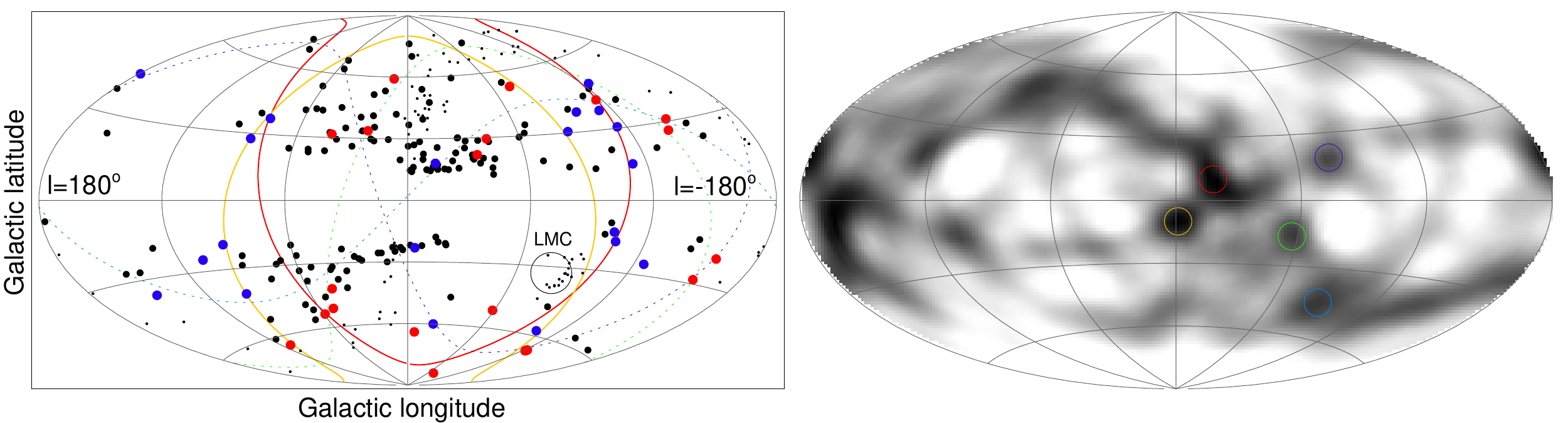}
    \caption[width=\textwidth]{{\it Left:} Distribution of the CRTS
      Mira stars with $P<170$d. Small black dots show the locations of
      stars in the Sgr stream and the LMC. Filled black circles
      correspond to the stars with $R<25$\,kpc, while filled blue
      circles to the stars with $25<R<50$\,kpc, and finally, red ones
      are those with $R>50$\,kpc. Red and yellow solid curves are the
      great circles corresponding to the two most significant
      over-densities in the plane of poles (see the Right
      panel). These great circles pass close to the LMC (marked with a
      large empty black circle) and are aligned with the direction of
      the dwarf's motion. Dotted green, blue and violet show the great
      circles corresponding to the three other, less significant
      over-densities in the plane of poles. {\it Right:} Great circle
      pole density. The grey-scale value of each pixel of this map
      corresponds to the density of stars along the great circle with
      the pole at this pixel. Stars are summed up with Gaussian
      weights according to their latitudinal distance from the equator
      of the great-circle coordinate frame with
      $\sigma=6^{\circ}$. The two most obvious over-densities are
      marked with red and yellow. Other, less significant candidate
      over-densities are marked with green, blue and violet.}
    \label{fig:short_period}
\end{figure*}
\begin{figure}
\centering
	\includegraphics[width=\columnwidth]{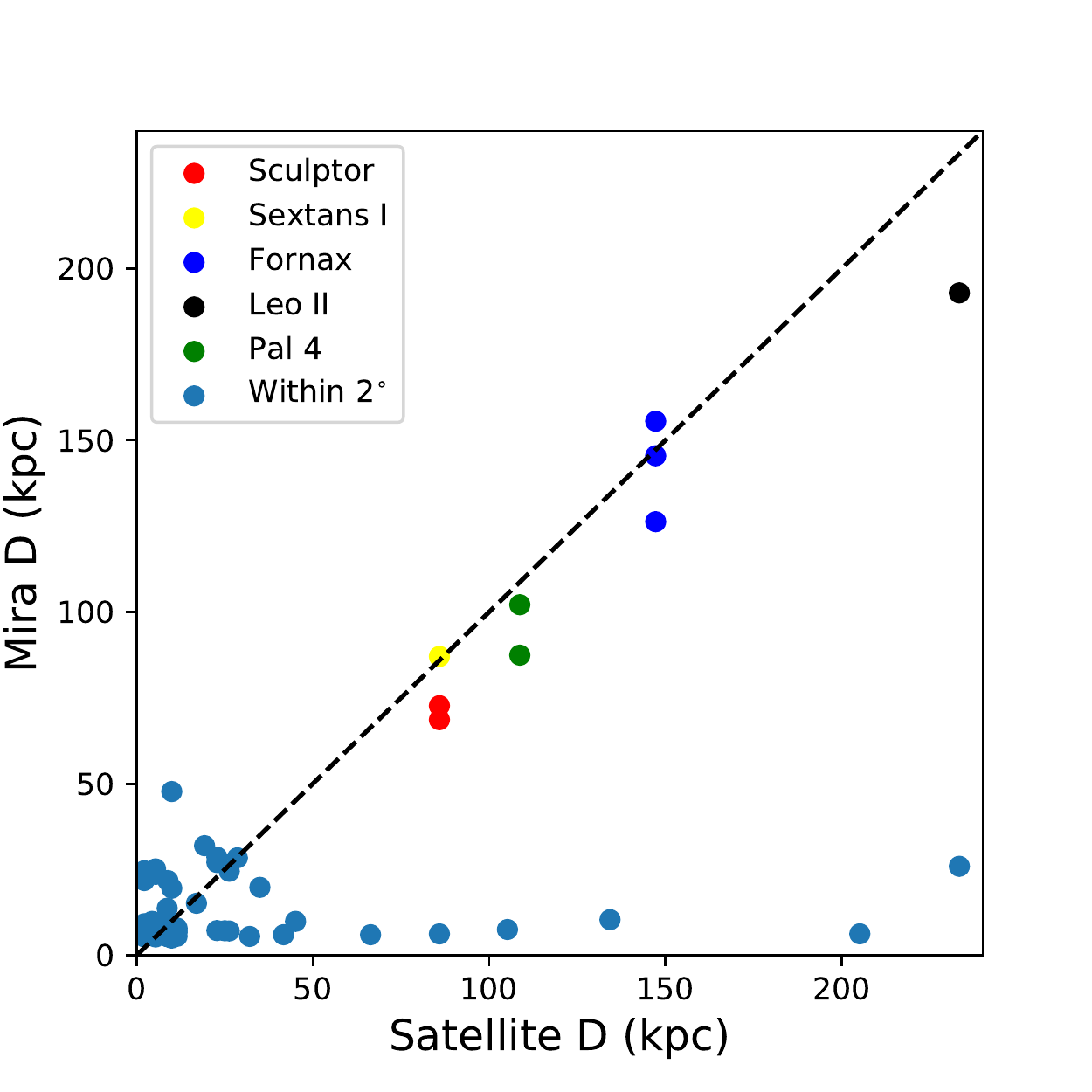}
    \caption[width=\columnwidth]{Possible associations of O-Mira with Local Group globular cluster and dwarf satellites. Red markers  show Mira within 0.5$^{\circ}$ of known satellites and blue markers within 2$^{\circ}$. The heliocentric distance offset from satellites is shown as deviation from the linear line. These Mira derive from the CRTS subset.}
    \label{fig:sats}
\end{figure}
The latest tally of the record holders can be found in
\citet{Bochanski2014}. Table 1 of the paper lists all of the 9 stars
with distances greater than 120 kpc known at the time. Of all
large-distance candidates, the most securely identified are the
horizontal branch stars. At least three independent samples of
pulsating HB stars, called RR Lyrae, have been published recently,
each containing sizeable numbers of stars beyond 100\,kpc
\citep[see][]{Drake2013,Sesar2017,Medina2018}. Additionally, blue HB
stars can also be used as standard candles -- albeit with somewhat
lower accuracy - and thus used to trace the outskirts of the
Galaxy. For example, \citet{Deason2012} uses a combination of deep
multi-band photometry and VLT spectroscopy to detect BHBs at distances
of order of 150 kpc. Finally, just like Miras, other stars around the
AGB are good candidates for the most distant objects in the Galaxy
owing to their spectacular luminosity. Accordingly, in the last two
decades, Carbon stars and M-giants have been used routinely to scout
the far reaches of the Milky Way
\citep[e.g.][]{Totten1998,Majewski2003,Mauron2008,Deason2012,Bochanski2014,Koposov2015,Mauron2018}.

Curiously, not only the four Southern (Galactic hemisphere) CRTS Miras
surpass all current distance records, they appear to cluster
conspicuously in a relatively small area of the sky. The group sits
around $l=345^{\circ} \pm 10^{\circ}$ and $b=-40^{\circ} \pm
17^{\circ}$. This portion of the celestial sphere is squeezed
inbetween the trailing tail of the Sgr dwarf and the trailing tail of
the LMC, with the four distant Mira positioned $23^{\circ} \pm
8^{\circ}$ away from the Sgr stream and $37^{\circ} \pm 14^{\circ}$
from the Magellanic Stream, in the coordinate system defined by
\citet{Nidever2008}. Even though the Miras are not aligned with either
of the streams perfectly, note that both the Sgr and the LMC debris
are predicted to travel to large distances around this location
\citep[][]{Dierickx2017,Diaz2012,Besla2013}. Most recently,
\citet{Deason2018hsc} have detected an excess of distant BHB stars in the ultra-deep data from the Subaru's Hyper Suprime-CAM at the location
not too far from that traced by the four Southern CRTS Miras. The
distant BHB over-density also overlaps with the so-called Pisces Cloud
\citep[][]{Sesar2007,Watkins2009,Nie2015}. While plenty of evidence
now exists that both dwarfs would leave trails of very distant stars
as their orbits evolve under the influence of the dynamical friction
\citep[see also][]{Gibbons2017,Jethwa2016}, it is less clear where on
the sky these debris should pile up. The uncertainty here is due to
the lack of constraints on the shape of the outer dark matter halo.

The whereabouts of the old Miras, i.e. those with short period
($P<170$\,d) can be studied in Fig.~\ref{fig:short_period_xyz}. Their
distribution appears to be broken into two components, a relatively
smooth, compact one with $r<20$\,kpc (as indicated by black circle)
and sparse lumpy one with $r>20$\,kpc. To investigate further the
spatial properties of the distant old Miras, we break them into three
sub-samples according to their Galactocentric distance, $r$ as shown in
Fig.~\ref{fig:short_period} in Galactic coordinates. Additionally,
for all short-period Miras with $r>25$\,kpc we carry out a great
circle star count, following ideas of
\citep[e.g.][]{Donald1995,Johnston1996,Ibata2001,Mateu2011}. In
practice, for a given pole, we assign a Gaussian (with
$\sigma=6^{\circ}$) weight to each Mira according to its latitude in
the great circle coordinate system. The density of stars in the plane
of poles is shown in the right panel of Fig.~\ref{fig:short_period}
in Galactic coordinates. The two most prominent over-densities are located
near $(l,b)=(0^{\circ}, 0^{\circ})$ and are marked with red and yellow
empty circles. The corresponding great circles are over-plotted in the
left panel. Interestingly, the two great circle pass
very close to the LMC and are approximately aligned with the motion of
the Cloud on the plane of the sky. This tentative detection of the
Magellanic debris traced by Miras is in agreement with the earlier study of 
\citet{De17}. Additionally, we identify three other, lower
significance over-densities marked in green, blue and violet, whose
corresponding great circles are shown in the left panel with dotted
lines.

\subsection{Satellite Galaxy and Globular Cluster Associations}

Although AGB stars have been surveyed in the Local Group dwarf spheroidals (dSphs), the only confirmed detection of an O-Mira is in Sextans~\citep{Sa12}. There are two C-Miras, but no O-Mira, so far found in Sculptor~\citep{Me11}. \citet{Wh09} report 6 C-Miras in Fornax -- a further candidate (F51010 in their notation) is so blue that they were unable to determine from its colours whether it was C-rich or O-rich. 

The spatial coordinates of the O-Mira are compared against those of known objects within the Local Group. We search for associations by requiring the O-Mira to lie less than half a degree from the centers of globular clusters and dSphs in the Local Group. We use half a degree because many of these objects are expected to have extra-tidal stars. Given that Miras correspond to a short-lived phase of stellar evolution, they are intrinsically extremely rare and so any chance association is a priori very unlikely. 

Potential associations exist with the Sculptor, Fornax, Sextans and the Leo II dwarf dSphs based on angular separation, as illustrated in Fig.~\ref{fig:sats}. Our candidates are checked to be O-Mira by confirming their location in Fig.~\ref{fig:cuts}. This shows all the candidate associations to be clearly O-Mira lying comfortably within our criteria.  The CRTS IDs, distances, periods and amplitudes of our candidates are listed in Table~\ref{table:data}. It would be interesting to confirm the candidates with kinematic evidence, which may already be possible with proper motion data from the forthcoming data release 2 of the Gaia satellite.

We also identified two O-Mira that are located close to the Palomar 4 (Pal 4) globular cluster. This is a young halo globular cluster that is likely to have been accreted by the Milky Way via the infall of its parent dwarf satellite. The two O-Mira are at heliocentric distances of 87 kpc and 102 kpc, compared to a distance of Pal 4 of 109 kpc~\citep{So03}. The latter authors carried out deep photometric searches in a field $1.3^\circ\times 0.9^\circ$ around Pal 4, and claimed evidence for the existence of extra-tidal tails and features. Although semi-regular variables are known in Pal 4, this is the first report of possible Miras.

\begin{table*}
\centering
\begin{tabular}{|cccccccc|}
\hline
\rowcolor[HTML]{C0C0C0} 
\textbf{CRTS ID} & \textbf{Period (days)} & \textbf{Amplitude} & \textbf{Mira D (kpc)} & \textbf{Satellite} & \textbf{Satellite D (kpc)} & {\color[HTML]{333333} \textbf{\begin{tabular}[c]{@{}c@{}}Age \\ estimate  \\ (Gyr)\end{tabular}}} & \textbf{\begin{tabular}[c]{@{}c@{}}Angular \\ offset  \\ (deg)\end{tabular}} \\
\hline
J005958.9-332834 & 196 & 1.1 & 72 $\pm$ 6 & Sculptor & 86 $\pm$ 6 & 6 & 0.2 \\
J010120.8-335304 & 92 & 0.8 & 69 $\pm$ 5 & Sculptor & 86 $\pm$ 6 & 10 & 0.3 \\
J101234.2-013440 & 120 & 1.1 & 87 $\pm$ 7 & Sextans (I) & 86 $\pm$ 4 & 10 & 0.1 \\
J023822.6-343804 & 163 & 0.8 & 145 $\pm$ 12 & Fornax & 147 $\pm$ 12 & 9 & 0.4 \\
J023910.7-343920 & 157 & 0.9 & 126 $\pm$ 10 & Fornax & 147 $\pm$ 12 & 10 & 0.3 \\
J024045.9-343453 & 143 & 0.8 & 156 $\pm$ 12 & Fornax & 147 $\pm$ 12 & 10 & 0.2 \\
J111320.6+221116 & 184 & 0.8 & 193 $\pm$ 15 & LeoII & 233 $\pm$ 14 & 7 & 0.05 \\
J112912.3+285815 & 151 & 1.9 & 87 $\pm$ 7 & Pal 4 & 100 & 10 & 0.02 \\
J112914.6+285814 & 131 & 1.3 & 102 $\pm$ 8 & Pal 4 & 100 & 10 & 0.009 \\
\hline
\end{tabular}

\caption{The data on the candidate O-Mira associations with Galactic dwarf spheroidals and distant globular clusters. We provide the CRTS ID and heliocentric coordinates of the O-Miras, as well as heliocentric distance to the satellite and angular offset.. Pal 4 distance estimates were obtained from \citet{Pal4_dist} and the remaining satellites from \citet{dSph_dist}. Tentative age estimates derive from the relation in Fig.~\ref{fig:period_age} and have been capped to 10 Gyr. Coordinates RA and Dec are encoded in the CRTS ID column and Mira distance errors derive similarly to those in Table~\ref{table:distant_mira}}
\label{table:data}
\end{table*}

\section{Conclusions}
\label{sec:conc}

This paper has highlighted a neglected resource. The Miras are bright
and can be traced throughout the Galaxy, and beyond. The carbon-rich
or C-Miras have a long history of use as tracers of halo
substructure~\citep[e.g.,][]{Do04,Hu15}. However, the oxygen-rich or
O-Miras have been less widely scrutinised, yet they are at least as
valuable. Their period and luminosity only weakly depends on colour,
so the O-Miras are excellent distance indicators. Their period depends
on age, with young O-Miras having the longest periods, so the O-Miras
are excellent chronometers. So, samples of Miras will enable us to
detect age gradients throughout the Galaxy and to date the Galactic
populations. Here, we have used the Catalina Surveys
and the All Sky Automated Survey for Supernovae to extract a clean
sample of $\sim 2,400$ O-Miras, the largest assembled to date. They
probe the Galactic bulge, thin and thick discs and the distant
halo. In principle, they offer a unique resource to age-date all the
structural components of the Galaxy.

We show that the disc O-Miras have periods increasing on moving
outwards from $\sim 3$ to 15 kpc. As the period of O-Miras correlates
inversely with age, this is a clear demonstration of the 'inside out'
nature of the Galactic disc~\citep[e.g.,][]{RG_gradient}. The outer
disc O-Miras have median period $\sim 275$ days (roughly 1-3 Gyrs), as
compared to the inner disc value of $\sim 200$ days (roughly 8-10
Gyrs). At least out to 15 kpc, there is no evidence of an upturn in
the radial profile of the age of disc, as expected from models in
which the outer disc is populated by old stars which migrated from the
inner disc~\citep{Ro08}. There are also hints of vertical variation,
in the sense that O-Miras at higher $|Z|$ comprise an older population
for given radius $R$. At smaller Galactocentric distances, within 2-3
kpc from the MW center, age gradients flatten out, signaling that the
bulge contains a mix of stellar ages, in agreement with earlier
studies \citep[see e.g.][]{Catchpole_ages}. Moving away from the
center, clearly visible in our data is the transition from younger
disc to older halo O-Miras, which occurs at $r \sim 15$ kpc where the
median O-Mira period plummets. The median period of the halo O-Miras
is $\sim 150$ days (roughly 10 Gyrs).

O-Mira are also precious as tracers outside the Galactic disc. The
tails from the Sagittarius are detectable in O-Miras, much like their
brethren the C-Miras. Moreover, we find strong evidence that for the
O-Miras in the Magellanic Stream at large distances from the
LMC. There is also a population of long period, and hence young,
O-Miras that follow the extent of the Galactic disc, but lie at
heights of up to 10 kpc. These may be stars kicked out of the disc,
perhaps analogous to other high latitude
structures~\citep{Bergemann2018,Sheffield2018,Deason2018}. The O-Miras
are so intrinsically luminous that they can trace the very outermost
reaches of the halo. We identify seven long-period (and therefore
likely young) O-Miras reaching distances between 200 and 500 kpc, much
beyond presently known record holders. A good fraction of the Miras
with extreme distances appear clustered on the sky, not too far from
the position where distant extensions of the Sgr stream, the LMC
trailing arm and the Pisces Cloud are expected. In addition to the
detection of distant debris piles associated with accretion events, we
have also provided new detections of O-Miras in Milky Way
satellites. Specifically, we found associations between O-Mira and the
Fornax, Sculptor, Sextans and Leo II Galactic dwarf spheroidals, as
well as the distant globular cluster Pal 4.

The Gaia satellite will dramatically increase the data on Mira. Although the main variable star data release occurs in 2020, Gaia Data Release 2 will already provide classifications for $\sim$ 500,000 variable stars, including many Miras. This offers the opportunity to select high quality samples of O-Miras, use the period-age relationship as a prior and provide ages of the main Galactic components as a function of distance. This will provide independent confirmation of ages estimates derived from isochrones. The opportunity to slice the Galaxy by age will soon be given to us. We are optimistic that the future for Miras is very bright.
 
\section*{Acknowledgments}

JG acknowledges financial support from the Science and Technology
Facilities Council. We thank members of the Cambridge Streams group
for helpful comments as this work was in progress. Iulia Simion, Alis
Deason, Victor Debattista and Patricia Whitelock are thanked for
helpful comments on the draft manuscript. The research leading to
these results has received funding from the European Research Council
under the European Union's Seventh Framework Programme (FP/2007-2013)
/ ERC Grant Agreement n. 308024.




\bibliographystyle{mnras}
\bibliography{bibliography} 

\begin{thebibliography}{}
\makeatletter
\relax
\def\mn@urlcharsother{\let\do\@makeother \do\$\do\&\do\#\do\^\do\_\do\%\do\~}
\def\mn@doi{\begingroup\mn@urlcharsother \@ifnextchar [ {\mn@doi@}
  {\mn@doi@[]}}
\def\mn@doi@[#1]#2{\def\@tempa{#1}\ifx\@tempa\@empty \href
  {http://dx.doi.org/#2} {doi:#2}\else \href {http://dx.doi.org/#2} {#1}\fi
  \endgroup}
\def\mn@eprint#1#2{\mn@eprint@#1:#2::\@nil}
\def\mn@eprint@arXiv#1{\href {http://arxiv.org/abs/#1} {{\tt arXiv:#1}}}
\def\mn@eprint@dblp#1{\href {http://dblp.uni-trier.de/rec/bibtex/#1.xml}
  {dblp:#1}}
\def\mn@eprint@#1:#2:#3:#4\@nil{\def\@tempa {#1}\def\@tempb {#2}\def\@tempc
  {#3}\ifx \@tempc \@empty \let \@tempc \@tempb \let \@tempb \@tempa \fi \ifx
  \@tempb \@empty \def\@tempb {arXiv}\fi \@ifundefined
  {mn@eprint@\@tempb}{\@tempb:\@tempc}{\expandafter \expandafter \csname
  mn@eprint@\@tempb\endcsname \expandafter{\@tempc}}}

\bibitem[\protect\citeauthoryear{{Am{\^o}res}, {Robin}  \&
  {Reyl{\'e}}}{{Am{\^o}res} et~al.}{2017}]{Am17}
{Am{\^o}res} E.~B.,  {Robin} A.~C.,   {Reyl{\'e}} C.,  2017, \mn@doi [\aap]
  {10.1051/0004-6361/201628461}, \href
  {http://adsabs.harvard.edu/abs/2017A%26A...602A..67A} {602, A67}

\bibitem[\protect\citeauthoryear{{An} et~al.,}{{An} et~al.}{2004}]{An04}
{An} J.~H.,  et~al., 2004, \mn@doi [\mnras] {10.1111/j.1365-2966.2004.07853.x},
  \href {http://adsabs.harvard.edu/abs/2004MNRAS.351.1071A} {351, 1071}

\bibitem[\protect\citeauthoryear{{Battinelli} \& {Demers}}{{Battinelli} \&
  {Demers}}{2014}]{Ba14}
{Battinelli} P.,  {Demers} S.,  2014, \mn@doi [\aap]
  {10.1051/0004-6361/201423900}, \href
  {http://adsabs.harvard.edu/abs/2014A%26A...568A.100B} {568, A100}

\bibitem[\protect\citeauthoryear{{Baumgardt}, {Parmentier}, {Anders}  \&
  {Grebel}}{{Baumgardt} et~al.}{2013}]{Baumgardt2013}
{Baumgardt} H.,  {Parmentier} G.,  {Anders} P.,   {Grebel} E.~K.,  2013,
  \mn@doi [\mnras] {10.1093/mnras/sts667}, \href
  {https://ui.adsabs.harvard.edu/#abs/2013MNRAS.430..676B} {430, 676}

\bibitem[\protect\citeauthoryear{{Belokurov} et~al.,}{{Belokurov}
  et~al.}{2014}]{vasily_stream}
{Belokurov} V.,  et~al., 2014, \mn@doi [\mnras] {10.1093/mnras/stt1862}, \href
  {http://adsabs.harvard.edu/abs/2014MNRAS.437..116B} {437, 116}

\bibitem[\protect\citeauthoryear{{Bergemann} et~al.,}{{Bergemann}
  et~al.}{2018}]{Bergemann2018}
{Bergemann} M.,  et~al., 2018, \mn@doi [\nat] {10.1038/nature25490}, \href
  {http://adsabs.harvard.edu/abs/2018Natur.555..334B} {555, 334}

\bibitem[\protect\citeauthoryear{{Besla}, {Hernquist}  \& {Loeb}}{{Besla}
  et~al.}{2013}]{Besla2013}
{Besla} G.,  {Hernquist} L.,   {Loeb} A.,  2013, \mn@doi [\mnras]
  {10.1093/mnras/sts192}, \href
  {http://adsabs.harvard.edu/abs/2013MNRAS.428.2342B} {428, 2342}

\bibitem[\protect\citeauthoryear{{Blommaert} \& {Groenewegen}}{{Blommaert} \&
  {Groenewegen}}{2007}]{bulgemira_feast}
{Blommaert} J.~A.~D.~L.,  {Groenewegen} M.~A.~T.,  2007, in {Vallenari} A.,
  {Tantalo} R.,  {Portinari} L.,   {Moretti} A.,  eds,  Astronomical Society of
  the Pacific Conference Series Vol. 374, From Stars to Galaxies: Building the
  Pieces to Build Up the Universe. p.~193

\bibitem[\protect\citeauthoryear{{Bochanski}, {Willman}, {Caldwell},
  {Sanderson}, {West}, {Strader}  \& {Brown}}{{Bochanski}
  et~al.}{2014}]{Bochanski2014}
{Bochanski} J.~J.,  {Willman} B.,  {Caldwell} N.,  {Sanderson} R.,  {West}
  A.~A.,  {Strader} J.,   {Brown} W.,  2014, \mn@doi [\apjl]
  {10.1088/2041-8205/790/1/L5}, \href
  {http://adsabs.harvard.edu/abs/2014ApJ...790L...5B} {790, L5}

\bibitem[\protect\citeauthoryear{{Bono}, {Marconi}, {Cassisi}, {Caputo},
  {Gieren}  \& {Pietrzynski}}{{Bono} et~al.}{2005}]{Bo05}
{Bono} G.,  {Marconi} M.,  {Cassisi} S.,  {Caputo} F.,  {Gieren} W.,
  {Pietrzynski} G.,  2005, \mn@doi [\apj] {10.1086/427744}, \href
  {http://adsabs.harvard.edu/abs/2005ApJ...621..966B} {621, 966}

\bibitem[\protect\citeauthoryear{{Carollo} et~al.,}{{Carollo}
  et~al.}{2016}]{Carollo2016}
{Carollo} D.,  et~al., 2016, \mn@doi [Nature Physics] {10.1038/nphys3874},
  \href {http://adsabs.harvard.edu/abs/2016NatPh..12.1170C} {12, 1170}

\bibitem[\protect\citeauthoryear{{Carpenter}}{{Carpenter}}{2001}]{colour_transform}
{Carpenter} J.~M.,  2001, \mn@doi [\aj] {10.1086/320383}, \href
  {http://adsabs.harvard.edu/abs/2001AJ....121.2851C} {121, 2851}

\bibitem[\protect\citeauthoryear{{Catchpole}, {Whitelock}, {Feast}, {Hughes},
  {Irwin}  \& {Alard}}{{Catchpole} et~al.}{2016}]{Catchpole_ages}
{Catchpole} R.~M.,  {Whitelock} P.~A.,  {Feast} M.~W.,  {Hughes} S.~M.~G.,
  {Irwin} M.,   {Alard} C.,  2016, \mn@doi [\mnras] {10.1093/mnras/stv2372},
  \href {http://adsabs.harvard.edu/abs/2016MNRAS.455.2216C} {455, 2216}

\bibitem[\protect\citeauthoryear{{Cioni}, {Marquette}, {Loup}, {Azzopardi},
  {Habing}, {Lasserre}  \& {Lesquoy}}{{Cioni} et~al.}{2001}]{Cioni2001}
{Cioni} M.-R.~L.,  {Marquette} J.-B.,  {Loup} C.,  {Azzopardi} M.,  {Habing}
  H.~J.,  {Lasserre} T.,   {Lesquoy} E.,  2001, \mn@doi [\aap]
  {10.1051/0004-6361:20011143}, \href
  {http://adsabs.harvard.edu/abs/2001A%26A...377..945C} {377, 945}

\bibitem[\protect\citeauthoryear{{Deason} et~al.,}{{Deason}
  et~al.}{2012}]{Deason2012}
{Deason} A.~J.,  et~al., 2012, \mn@doi [\mnras]
  {10.1111/j.1365-2966.2012.21639.x}, \href
  {http://adsabs.harvard.edu/abs/2012MNRAS.425.2840D} {425, 2840}

\bibitem[\protect\citeauthoryear{{Deason}, {Belokurov}, {Erkal}, {Koposov}  \&
  {Mackey}}{{Deason} et~al.}{2017}]{De17}
{Deason} A.~J.,  {Belokurov} V.,  {Erkal} D.,  {Koposov} S.~E.,   {Mackey} D.,
  2017, \mn@doi [\mnras] {10.1093/mnras/stx263}, \href
  {http://adsabs.harvard.edu/abs/2017MNRAS.467.2636D} {467, 2636}

\bibitem[\protect\citeauthoryear{{Deason}, {Belokurov}  \& {Koposov}}{{Deason}
  et~al.}{2018a}]{Deason2018}
{Deason} A.~J.,  {Belokurov} V.,   {Koposov} S.~E.,  2018a, \mn@doi [\mnras]
  {10.1093/mnras/stx2528}, \href
  {http://adsabs.harvard.edu/abs/2018MNRAS.473.2428D} {473, 2428}

\bibitem[\protect\citeauthoryear{{Deason}, {Belokurov}  \& {Koposov}}{{Deason}
  et~al.}{2018b}]{Deason2018hsc}
{Deason} A.~J.,  {Belokurov} V.,   {Koposov} S.~E.,  2018b, \mn@doi [\apj]
  {10.3847/1538-4357/aa9d19}, \href
  {http://adsabs.harvard.edu/abs/2018ApJ...852..118D} {852, 118}

\bibitem[\protect\citeauthoryear{{Diaz} \& {Bekki}}{{Diaz} \&
  {Bekki}}{2012}]{Diaz2012}
{Diaz} J.~D.,  {Bekki} K.,  2012, \mn@doi [\apj] {10.1088/0004-637X/750/1/36},
  \href {http://adsabs.harvard.edu/abs/2012ApJ...750...36D} {750, 36}

\bibitem[\protect\citeauthoryear{{Dierickx} \& {Loeb}}{{Dierickx} \&
  {Loeb}}{2017}]{Dierickx2017}
{Dierickx} M.~I.~P.,  {Loeb} A.,  2017, \mn@doi [\apj]
  {10.3847/1538-4357/836/1/92}, \href
  {http://adsabs.harvard.edu/abs/2017ApJ...836...92D} {836, 92}

\bibitem[\protect\citeauthoryear{{Downes} et~al.,}{{Downes}
  et~al.}{2004}]{Do04}
{Downes} R.~A.,  et~al., 2004, \mn@doi [\aj] {10.1086/383211}, \href
  {http://adsabs.harvard.edu/abs/2004AJ....127.2838D} {127, 2838}

\bibitem[\protect\citeauthoryear{{Drake} et~al.,}{{Drake}
  et~al.}{2013}]{Drake2013}
{Drake} A.~J.,  et~al., 2013, \mn@doi [\apj] {10.1088/0004-637X/765/2/154},
  \href {http://adsabs.harvard.edu/abs/2013ApJ...765..154D} {765, 154}

\bibitem[\protect\citeauthoryear{{Drake} et~al.,}{{Drake}
  et~al.}{2014}]{crts_north}
{Drake} A.~J.,  et~al., 2014, \mn@doi [\apjs] {10.1088/0067-0049/213/1/9},
  \href {http://adsabs.harvard.edu/abs/2014ApJS..213....9D} {213, 9}

\bibitem[\protect\citeauthoryear{Drake et~al.,}{Drake
  et~al.}{2017}]{crts_south}
Drake A.~J.,  et~al., 2017, \mn@doi [Monthly Notices of the Royal Astronomical
  Society] {10.1093/mnras/stx1085}, 469, 3688

\bibitem[\protect\citeauthoryear{{Efremov}}{{Efremov}}{1978}]{Ef78}
{Efremov} I.~N.,  1978, \sovast, \href
  {http://adsabs.harvard.edu/abs/1978SvA....22..161E} {22, 161}

\bibitem[\protect\citeauthoryear{{Elgueta} et~al.,}{{Elgueta}
  et~al.}{2016}]{lmc_dist1}
{Elgueta} S.~S.,  et~al., 2016, \mn@doi [\aj] {10.3847/0004-6256/152/2/29},
  \href {http://adsabs.harvard.edu/abs/2016AJ....152...29E} {152, 29}

\bibitem[\protect\citeauthoryear{{Feast}}{{Feast}}{2007}]{Fe07}
{Feast} M.,  2007, in {Kerschbaum} F.,  {Charbonnel} C.,   {Wing} R.~F.,  eds,
  Astronomical Society of the Pacific Conference Series Vol. 378, Why Galaxies
  Care About AGB Stars: Their Importance as Actors and Probes. p.~479
  (\mn@eprint {} {astro-ph/0609318})

\bibitem[\protect\citeauthoryear{{Feast}}{{Feast}}{2009}]{Fe09}
{Feast} M.~W.,  2009, in {Ueta} T.,  {Matsunaga} N.,   {Ita} Y.,  eds, AGB
  Stars and Related Phenomena. p.~48 (\mn@eprint {arXiv} {0812.0250})

\bibitem[\protect\citeauthoryear{{Feast} \& {Whitelock}}{{Feast} \&
  {Whitelock}}{2000}]{Fe00}
{Feast} M.~W.,  {Whitelock} P.~A.,  2000, \mn@doi [\mnras]
  {10.1046/j.1365-8711.2000.03629.x}, \href
  {http://adsabs.harvard.edu/abs/2000MNRAS.317..460F} {317, 460}

\bibitem[\protect\citeauthoryear{{Feast}, {Glass}, {Whitelock}  \&
  {Catchpole}}{{Feast} et~al.}{1989}]{Fe89}
{Feast} M.~W.,  {Glass} I.~S.,  {Whitelock} P.~A.,   {Catchpole} R.~M.,  1989,
  \mn@doi [\mnras] {10.1093/mnras/241.3.375}, \href
  {http://adsabs.harvard.edu/abs/1989MNRAS.241..375F} {241, 375}

\bibitem[\protect\citeauthoryear{{Feast}, {Whitelock}  \& {Menzies}}{{Feast}
  et~al.}{2006}]{Fe06}
{Feast} M.~W.,  {Whitelock} P.~A.,   {Menzies} J.~W.,  2006, \mn@doi [\mnras]
  {10.1111/j.1365-2966.2006.10324.x}, \href
  {http://adsabs.harvard.edu/abs/2006MNRAS.369..791F} {369, 791}

\bibitem[\protect\citeauthoryear{{Gibbons}, {Belokurov}  \& {Evans}}{{Gibbons}
  et~al.}{2017}]{Gibbons2017}
{Gibbons} S.~L.~J.,  {Belokurov} V.,   {Evans} N.~W.,  2017, \mn@doi [\mnras]
  {10.1093/mnras/stw2328}, \href
  {http://adsabs.harvard.edu/abs/2017MNRAS.464..794G} {464, 794}

\bibitem[\protect\citeauthoryear{{Glass}, {Whitelock}, {Catchpole}  \&
  {Feast}}{{Glass} et~al.}{1995}]{glass_orich_line}
{Glass} I.~S.,  {Whitelock} P.~A.,  {Catchpole} R.~M.,   {Feast} M.~W.,  1995,
  \mn@doi [\mnras] {10.1093/mnras/273.2.383}, \href
  {http://adsabs.harvard.edu/abs/1995MNRAS.273..383G} {273, 383}

\bibitem[\protect\citeauthoryear{{Habing} \& {Olofsson}}{{Habing} \&
  {Olofsson}}{2003}]{Ha03}
{Habing} H.~J.,  {Olofsson} H.,  eds, 2003, {Asymptotic giant branch stars}

\bibitem[\protect\citeauthoryear{{Harris}}{{Harris}}{1996}]{Pal4_dist}
{Harris} W.~E.,  1996, \mn@doi [\aj] {10.1086/118116}, \href
  {http://adsabs.harvard.edu/abs/1996AJ....112.1487H} {112, 1487}

\bibitem[\protect\citeauthoryear{{Huang} et~al.,}{{Huang} et~al.}{2018}]{Hu18}
{Huang} C.~D.,  et~al., 2018, preprint, \href
  {http://adsabs.harvard.edu/abs/2018arXiv180102711H} {} (\mn@eprint {arXiv}
  {1801.02711})

\bibitem[\protect\citeauthoryear{{Huxor} \& {Grebel}}{{Huxor} \&
  {Grebel}}{2015}]{Hu15}
{Huxor} A.~P.,  {Grebel} E.~K.,  2015, \mn@doi [\mnras]
  {10.1093/mnras/stv1631}, \href
  {http://adsabs.harvard.edu/abs/2015MNRAS.453.2653H} {453, 2653}

\bibitem[\protect\citeauthoryear{{Ibata}, {Lewis}, {Irwin}, {Totten}  \&
  {Quinn}}{{Ibata} et~al.}{2001}]{Ibata2001}
{Ibata} R.,  {Lewis} G.~F.,  {Irwin} M.,  {Totten} E.,   {Quinn} T.,  2001,
  \mn@doi [\apj] {10.1086/320060}, \href
  {http://adsabs.harvard.edu/abs/2001ApJ...551..294I} {551, 294}

\bibitem[\protect\citeauthoryear{{Jayaraman}, {Gilmore}, {Wyse}, {Norris}  \&
  {Belokurov}}{{Jayaraman} et~al.}{2013}]{thickdisk_jaya}
{Jayaraman} A.,  {Gilmore} G.,  {Wyse} R.~F.~G.,  {Norris} J.~E.,   {Belokurov}
  V.,  2013, \mn@doi [\mnras] {10.1093/mnras/stt221}, \href
  {http://adsabs.harvard.edu/abs/2013MNRAS.431..930J} {431, 930}

\bibitem[\protect\citeauthoryear{{Jayasinghe} et~al.,}{{Jayasinghe}
  et~al.}{2018}]{ASAS}
{Jayasinghe} T.,  et~al., 2018, preprint, \href
  {http://adsabs.harvard.edu/abs/2018arXiv180301001J} {} (\mn@eprint {arXiv}
  {1803.01001})

\bibitem[\protect\citeauthoryear{{Jethwa}, {Erkal}  \& {Belokurov}}{{Jethwa}
  et~al.}{2016}]{Jethwa2016}
{Jethwa} P.,  {Erkal} D.,   {Belokurov} V.,  2016, \mn@doi [\mnras]
  {10.1093/mnras/stw1343}, \href
  {http://adsabs.harvard.edu/abs/2016MNRAS.461.2212J} {461, 2212}

\bibitem[\protect\citeauthoryear{{Johnston}, {Hernquist}  \&
  {Bolte}}{{Johnston} et~al.}{1996}]{Johnston1996}
{Johnston} K.~V.,  {Hernquist} L.,   {Bolte} M.,  1996, \mn@doi [\apj]
  {10.1086/177418}, \href {http://adsabs.harvard.edu/abs/1996ApJ...465..278J}
  {465, 278}

\bibitem[\protect\citeauthoryear{{Juri{\'c}} et~al.,}{{Juri{\'c}}
  et~al.}{2008}]{thickdisk_juric}
{Juri{\'c}} M.,  et~al., 2008, \mn@doi [\apj] {10.1086/523619}, \href
  {http://adsabs.harvard.edu/abs/2008ApJ...673..864J} {673, 864}

\bibitem[\protect\citeauthoryear{{Kamath}, {Wood}, {Soszy{\'n}ski}  \&
  {Lebzelter}}{{Kamath} et~al.}{2010}]{Kamath2010}
{Kamath} D.,  {Wood} P.~R.,  {Soszy{\'n}ski} I.,   {Lebzelter} T.,  2010,
  \mn@doi [\mnras] {10.1111/j.1365-2966.2010.17137.x}, \href
  {http://adsabs.harvard.edu/abs/2010MNRAS.408..522K} {408, 522}

\bibitem[\protect\citeauthoryear{{Kharchenko}, {Piskunov}, {Schilbach},
  {R{\"o}ser}  \& {Scholz}}{{Kharchenko} et~al.}{2016}]{Kharchenko2016}
{Kharchenko} N.~V.,  {Piskunov} A.~E.,  {Schilbach} E.,  {R{\"o}ser} S.,
  {Scholz} R.~D.,  2016, \mn@doi [\aap] {10.1051/0004-6361/201527292}, \href
  {https://ui.adsabs.harvard.edu/#abs/2016A&A...585A.101K} {585}

\bibitem[\protect\citeauthoryear{{Koposov}, {Belokurov}, {Zucker}, {Lewis},
  {Ibata}, {Olszewski}, {L{\'o}pez-S{\'a}nchez}  \& {Hyde}}{{Koposov}
  et~al.}{2015}]{Koposov2015}
{Koposov} S.~E.,  {Belokurov} V.,  {Zucker} D.~B.,  {Lewis} G.~F.,  {Ibata}
  R.~A.,  {Olszewski} E.~W.,  {L{\'o}pez-S{\'a}nchez} {\'A}.~R.,   {Hyde}
  E.~A.,  2015, \mn@doi [\mnras] {10.1093/mnras/stu2263}, \href
  {http://adsabs.harvard.edu/abs/2015MNRAS.446.3110K} {446, 3110}

\bibitem[\protect\citeauthoryear{{L{\'o}pez-Corredoira}}{{L{\'o}pez-Corredoira}}{2017}]{mira_age}
{L{\'o}pez-Corredoira} M.,  2017, \mn@doi [\apj] {10.3847/1538-4357/836/2/218},
  \href {http://adsabs.harvard.edu/abs/2017ApJ...836..218L} {836, 218}

\bibitem[\protect\citeauthoryear{{Lorenz}, {Lebzelter}, {Nowotny}, {Telting},
  {Kerschbaum}, {Olofsson}  \& {Schwarz}}{{Lorenz} et~al.}{2011}]{Lorenz2011}
{Lorenz} D.,  {Lebzelter} T.,  {Nowotny} W.,  {Telting} J.,  {Kerschbaum} F.,
  {Olofsson} H.,   {Schwarz} H.~E.,  2011, \mn@doi [\aap]
  {10.1051/0004-6361/201116951}, \href
  {http://adsabs.harvard.edu/abs/2011A%26A...532A..78L} {532, A78}

\bibitem[\protect\citeauthoryear{{Lynden-Bell} \& {Lynden-Bell}}{{Lynden-Bell}
  \& {Lynden-Bell}}{1995}]{Donald1995}
{Lynden-Bell} D.,  {Lynden-Bell} R.~M.,  1995, \mn@doi [\mnras]
  {10.1093/mnras/275.2.429}, \href
  {http://adsabs.harvard.edu/abs/1995MNRAS.275..429L} {275, 429}

\bibitem[\protect\citeauthoryear{{Majewski}, {Skrutskie}, {Weinberg}  \&
  {Ostheimer}}{{Majewski} et~al.}{2003}]{Majewski2003}
{Majewski} S.~R.,  {Skrutskie} M.~F.,  {Weinberg} M.~D.,   {Ostheimer} J.~C.,
  2003, \mn@doi [\apj] {10.1086/379504}, \href
  {http://adsabs.harvard.edu/abs/2003ApJ...599.1082M} {599, 1082}

\bibitem[\protect\citeauthoryear{{Martig} et~al.,}{{Martig}
  et~al.}{2016a}]{Ma16}
{Martig} M.,  et~al., 2016a, \mn@doi [\mnras] {10.1093/mnras/stv2830}, \href
  {http://adsabs.harvard.edu/abs/2016MNRAS.456.3655M} {456, 3655}

\bibitem[\protect\citeauthoryear{{Martig}, {Minchev}, {Ness}, {Fouesneau}  \&
  {Rix}}{{Martig} et~al.}{2016b}]{RG_gradient}
{Martig} M.,  {Minchev} I.,  {Ness} M.,  {Fouesneau} M.,   {Rix} H.-W.,  2016b,
  \mn@doi [\apj] {10.3847/0004-637X/831/2/139}, \href
  {http://adsabs.harvard.edu/abs/2016ApJ...831..139M} {831, 139}

\bibitem[\protect\citeauthoryear{{Mateu}, {Bruzual}, {Aguilar}, {Brown},
  {Valenzuela}, {Carigi}, {Vel{\'a}zquez}  \& {Hern{\'a}ndez}}{{Mateu}
  et~al.}{2011}]{Mateu2011}
{Mateu} C.,  {Bruzual} G.,  {Aguilar} L.,  {Brown} A.~G.~A.,  {Valenzuela} O.,
  {Carigi} L.,  {Vel{\'a}zquez} H.,   {Hern{\'a}ndez} F.,  2011, \mn@doi
  [\mnras] {10.1111/j.1365-2966.2011.18690.x}, \href
  {http://adsabs.harvard.edu/abs/2011MNRAS.415..214M} {415, 214}

\bibitem[\protect\citeauthoryear{{Matsunaga}, {Menzies}, {Feast}, {Whitelock},
  {Onozato}, {Barway}  \& {Aydi}}{{Matsunaga} et~al.}{2017}]{bulgemira_mats}
{Matsunaga} N.,  {Menzies} J.~W.,  {Feast} M.~W.,  {Whitelock} P.~A.,
  {Onozato} H.,  {Barway} S.,   {Aydi} E.,  2017, \mn@doi [\mnras]
  {10.1093/mnras/stx1213}, \href
  {http://adsabs.harvard.edu/abs/2017MNRAS.469.4949M} {469, 4949}

\bibitem[\protect\citeauthoryear{{Matsuura} et~al.,}{{Matsuura}
  et~al.}{2009}]{Matsuura2009}
{Matsuura} M.,  et~al., 2009, \mn@doi [\mnras]
  {10.1111/j.1365-2966.2009.14743.x}, \href
  {http://adsabs.harvard.edu/abs/2009MNRAS.396..918M} {396, 918}

\bibitem[\protect\citeauthoryear{{Mauron}}{{Mauron}}{2008}]{Mauron2008}
{Mauron} N.,  2008, \mn@doi [\aap] {10.1051/0004-6361:200809378}, \href
  {http://adsabs.harvard.edu/abs/2008A%26A...482..151M} {482, 151}

\bibitem[\protect\citeauthoryear{{Mauron}, {Gigoyan}  \& {Kostandyan}}{{Mauron}
  et~al.}{2018}]{Mauron2018}
{Mauron} N.,  {Gigoyan} K.~S.,   {Kostandyan} G.~R.,  2018, \mn@doi
  [Astrophysics] {10.1007/s10511-018-9517-x}, \href
  {http://adsabs.harvard.edu/abs/2018Ap.....61...83M} {61, 83}

\bibitem[\protect\citeauthoryear{{McConnachie}}{{McConnachie}}{2012}]{dSph_dist}
{McConnachie} A.~W.,  2012, \mn@doi [\aj] {10.1088/0004-6256/144/1/4}, \href
  {http://adsabs.harvard.edu/abs/2012AJ....144....4M} {144, 4}

\bibitem[\protect\citeauthoryear{{Medina} et~al.,}{{Medina}
  et~al.}{2018}]{Medina2018}
{Medina} G.~E.,  et~al., 2018, \mn@doi [\apj] {10.3847/1538-4357/aaad02}, \href
  {http://adsabs.harvard.edu/abs/2018ApJ...855...43M} {855, 43}

\bibitem[\protect\citeauthoryear{{Menzies}, {Feast}, {Whitelock}  \&
  {Matsunaga}}{{Menzies} et~al.}{2011}]{Me11}
{Menzies} J.~W.,  {Feast} M.~W.,  {Whitelock} P.~A.,   {Matsunaga} N.,  2011,
  \mn@doi [\mnras] {10.1111/j.1365-2966.2011.18649.x}, \href
  {http://adsabs.harvard.edu/abs/2011MNRAS.414.3492M} {414, 3492}

\bibitem[\protect\citeauthoryear{{Ness}, {Hogg}, {Rix}, {Martig},
  {Pinsonneault}  \& {Ho}}{{Ness} et~al.}{2016}]{Ne16}
{Ness} M.,  {Hogg} D.~W.,  {Rix} H.-W.,  {Martig} M.,  {Pinsonneault} M.~H.,
  {Ho} A.~Y.~Q.,  2016, \mn@doi [\apj] {10.3847/0004-637X/823/2/114}, \href
  {http://adsabs.harvard.edu/abs/2016ApJ...823..114N} {823, 114}

\bibitem[\protect\citeauthoryear{{Nidever}, {Majewski}  \& {Butler
  Burton}}{{Nidever} et~al.}{2008}]{Nidever2008}
{Nidever} D.~L.,  {Majewski} S.~R.,   {Butler Burton} W.,  2008, \mn@doi [\apj]
  {10.1086/587042}, \href {http://adsabs.harvard.edu/abs/2008ApJ...679..432N}
  {679, 432}

\bibitem[\protect\citeauthoryear{{Nie} et~al.,}{{Nie} et~al.}{2015}]{Nie2015}
{Nie} J.~D.,  et~al., 2015, \mn@doi [\apj] {10.1088/0004-637X/810/2/153}, \href
  {http://adsabs.harvard.edu/abs/2015ApJ...810..153N} {810, 153}

\bibitem[\protect\citeauthoryear{{Nishida}, {Tanab{\'e}}, {Nakada},
  {Matsumoto}, {Sekiguchi}  \& {Glass}}{{Nishida} et~al.}{2000}]{Nishida2000}
{Nishida} S.,  {Tanab{\'e}} T.,  {Nakada} Y.,  {Matsumoto} S.,  {Sekiguchi} K.,
    {Glass} I.~S.,  2000, \mn@doi [\mnras] {10.1046/j.1365-8711.2000.03189.x},
  \href {http://adsabs.harvard.edu/abs/2000MNRAS.313..136N} {313, 136}

\bibitem[\protect\citeauthoryear{{Ojha}}{{Ojha}}{2001}]{thickdsk_ojha}
{Ojha} D.~K.,  2001, \mn@doi [\mnras] {10.1046/j.1365-8711.2001.04155.x}, \href
  {http://adsabs.harvard.edu/abs/2001MNRAS.322..426O} {322, 426}

\bibitem[\protect\citeauthoryear{{Rocha-Pinto}, {Majewski}, {Skrutskie},
  {Crane}  \& {Patterson}}{{Rocha-Pinto} et~al.}{2004}]{RochaPinto2004}
{Rocha-Pinto} H.~J.,  {Majewski} S.~R.,  {Skrutskie} M.~F.,  {Crane} J.~D.,
  {Patterson} R.~J.,  2004, \mn@doi [\apj] {10.1086/424585}, \href
  {http://adsabs.harvard.edu/abs/2004ApJ...615..732R} {615, 732}

\bibitem[\protect\citeauthoryear{{Ro{\v s}kar}, {Debattista}, {Stinson},
  {Quinn}, {Kaufmann}  \& {Wadsley}}{{Ro{\v s}kar} et~al.}{2008}]{Ro08}
{Ro{\v s}kar} R.,  {Debattista} V.~P.,  {Stinson} G.~S.,  {Quinn} T.~R.,
  {Kaufmann} T.,   {Wadsley} J.,  2008, \mn@doi [\apjl] {10.1086/586734}, \href
  {http://adsabs.harvard.edu/abs/2008ApJ...675L..65R} {675, L65}

\bibitem[\protect\citeauthoryear{{Sakamoto}, {Matsunaga}, {Hasegawa}  \&
  {Nakada}}{{Sakamoto} et~al.}{2012}]{Sa12}
{Sakamoto} T.,  {Matsunaga} N.,  {Hasegawa} T.,   {Nakada} Y.,  2012, \mn@doi
  [\apjl] {10.1088/2041-8205/761/1/L10}, \href
  {http://adsabs.harvard.edu/abs/2012ApJ...761L..10S} {761, L10}

\bibitem[\protect\citeauthoryear{{Schlegel}, {Finkbeiner}  \&
  {Davis}}{{Schlegel} et~al.}{1998}]{schlegel}
{Schlegel} D.~J.,  {Finkbeiner} D.~P.,   {Davis} M.,  1998, \mn@doi [\apj]
  {10.1086/305772}, \href {http://adsabs.harvard.edu/abs/1998ApJ...500..525S}
  {500, 525}

\bibitem[\protect\citeauthoryear{{Sesar} et~al.,}{{Sesar}
  et~al.}{2007}]{Sesar2007}
{Sesar} B.,  et~al., 2007, \mn@doi [\aj] {10.1086/521819}, \href
  {http://adsabs.harvard.edu/abs/2007AJ....134.2236S} {134, 2236}

\bibitem[\protect\citeauthoryear{{Sesar} et~al.,}{{Sesar}
  et~al.}{2017}]{Sesar2017}
{Sesar} B.,  et~al., 2017, \mn@doi [\aj] {10.3847/1538-3881/aa661b}, \href
  {http://adsabs.harvard.edu/abs/2017AJ....153..204S} {153, 204}

\bibitem[\protect\citeauthoryear{{Sheffield}, {Price-Whelan}, {Tzanidakis},
  {Johnston}, {Laporte}  \& {Sesar}}{{Sheffield} et~al.}{2018}]{Sheffield2018}
{Sheffield} A.~A.,  {Price-Whelan} A.~M.,  {Tzanidakis} A.,  {Johnston} K.~V.,
  {Laporte} C.~F.~P.,   {Sesar} B.,  2018, \mn@doi [\apj]
  {10.3847/1538-4357/aaa4b6}, \href
  {http://adsabs.harvard.edu/abs/2018ApJ...854...47S} {854, 47}

\bibitem[\protect\citeauthoryear{{Sohn} et~al.,}{{Sohn} et~al.}{2003}]{So03}
{Sohn} Y.-J.,  et~al., 2003, \mn@doi [\aj] {10.1086/375907}, \href
  {http://adsabs.harvard.edu/abs/2003AJ....126..803S} {126, 803}

\bibitem[\protect\citeauthoryear{{Soszy{\'n}ski} et~al.,}{{Soszy{\'n}ski}
  et~al.}{2009a}]{ogle_lpv_lmc}
{Soszy{\'n}ski} I.,  et~al., 2009a, \actaa, \href
  {http://adsabs.harvard.edu/abs/2009AcA....59..239S} {59, 239}

\bibitem[\protect\citeauthoryear{{Soszy{\'n}ski} et~al.,}{{Soszy{\'n}ski}
  et~al.}{2009b}]{Soszynski2009}
{Soszy{\'n}ski} I.,  et~al., 2009b, \actaa, \href
  {https://ui.adsabs.harvard.edu/#abs/2009AcA....59..239S} {59, 239}

\bibitem[\protect\citeauthoryear{{Totten} \& {Irwin}}{{Totten} \&
  {Irwin}}{1998}]{Totten1998}
{Totten} E.~J.,  {Irwin} M.~J.,  1998, \mn@doi [\mnras]
  {10.1046/j.1365-8711.1998.01086.x}, \href
  {http://adsabs.harvard.edu/abs/1998MNRAS.294....1T} {294, 1}

\bibitem[\protect\citeauthoryear{{Watkins} et~al.,}{{Watkins}
  et~al.}{2009}]{Watkins2009}
{Watkins} L.~L.,  et~al., 2009, \mn@doi [\mnras]
  {10.1111/j.1365-2966.2009.15242.x}, \href
  {http://adsabs.harvard.edu/abs/2009MNRAS.398.1757W} {398, 1757}

\bibitem[\protect\citeauthoryear{{Whitelock}, {Feast}, {Marang}  \&
  {Groenewegen}}{{Whitelock} et~al.}{2006}]{Whitelock2006}
{Whitelock} P.~A.,  {Feast} M.~W.,  {Marang} F.,   {Groenewegen} M.~A.~T.,
  2006, \mn@doi [\mnras] {10.1111/j.1365-2966.2006.10322.x}, \href
  {http://adsabs.harvard.edu/abs/2006MNRAS.369..751W} {369, 751}

\bibitem[\protect\citeauthoryear{{Whitelock}, {Feast}  \& {Van
  Leeuwen}}{{Whitelock} et~al.}{2008}]{Wh08}
{Whitelock} P.~A.,  {Feast} M.~W.,   {Van Leeuwen} F.,  2008, \mn@doi [\mnras]
  {10.1111/j.1365-2966.2008.13032.x}, \href
  {http://adsabs.harvard.edu/abs/2008MNRAS.386..313W} {386, 313}

\bibitem[\protect\citeauthoryear{{Whitelock}, {Menzies}, {Feast}, {Matsunaga},
  {Tanab{\'e}}  \& {Ita}}{{Whitelock} et~al.}{2009}]{Wh09}
{Whitelock} P.~A.,  {Menzies} J.~W.,  {Feast} M.~W.,  {Matsunaga} N.,
  {Tanab{\'e}} T.,   {Ita} Y.,  2009, \mn@doi [\mnras]
  {10.1111/j.1365-2966.2008.14365.x}, \href
  {http://adsabs.harvard.edu/abs/2009MNRAS.394..795W} {394, 795}

\bibitem[\protect\citeauthoryear{{Xu}, {Newberg}, {Carlin}, {Liu}, {Deng},
  {Li}, {Sch{\"o}nrich}  \& {Yanny}}{{Xu} et~al.}{2015}]{Xu2015}
{Xu} Y.,  {Newberg} H.~J.,  {Carlin} J.~L.,  {Liu} C.,  {Deng} L.,  {Li} J.,
  {Sch{\"o}nrich} R.,   {Yanny} B.,  2015, \mn@doi [\apj]
  {10.1088/0004-637X/801/2/105}, \href
  {http://adsabs.harvard.edu/abs/2015ApJ...801..105X} {801, 105}

\bibitem[\protect\citeauthoryear{{Yuan}, {Macri}, {He}, {Huang}, {Kanbur}  \&
  {Ngeow}}{{Yuan} et~al.}{2017}]{Yu17}
{Yuan} W.,  {Macri} L.~M.,  {He} S.,  {Huang} J.~Z.,  {Kanbur} S.~M.,   {Ngeow}
  C.-C.,  2017, \mn@doi [\aj] {10.3847/1538-3881/aa86f1}, \href
  {http://adsabs.harvard.edu/abs/2017AJ....154..149Y} {154, 149}

\bibitem[\protect\citeauthoryear{{de Boer}, {Belokurov}  \& {Koposov}}{{de
  Boer} et~al.}{2018}]{deBoer2018}
{de Boer} T.~J.~L.,  {Belokurov} V.,   {Koposov} S.~E.,  2018, \mn@doi [\mnras]
  {10.1093/mnras/stx2391}, \href
  {http://adsabs.harvard.edu/abs/2018MNRAS.473..647D} {473, 647}

\makeatother
\end{thebibliography}




\bsp	
\label{lastpage}
\end{document}